\documentclass[fleqn,usenatbib]{mnras}

\usepackage{ae,aecompl}
\usepackage{amsmath}	
\usepackage{amssymb}	
\usepackage{ulem}
\DeclareUnicodeCharacter{2212}{-}

\usepackage[T1]{fontenc}

\DeclareRobustCommand{\VAN}[3]{#2}
\let\VANthebibliography\thebibliography
\def\thebibliography{\DeclareRobustCommand{\VAN}[3]{##3}\VANthebibliography}

\usepackage{graphicx}	
\usepackage{amsmath}	
\usepackage{amssymb}	

\title[IFU observations of the inner 200 pc of NGC\,4546]{IFU observations of the inner 200 pc of NGC\,4546: gas rotation, non-circular motions and ionised outflows}

\author[K. F. Heckler et al.]{Kelly F. Heckler,$^{1}$\thanks{E-mail: ke.heckler95@gmail.com}
Tiago V. Ricci,$^{2}$
Rogemar A. Riffel$^{1}$
\\
$^{1}$ Departamento de F\'isica, Universidade Federal de Santa Maria, 97105-900 Santa Maria, RS, Brazil
\\
$^{2}$ Universidade Federal da Fronteira Sul, Campus Cerro Largo, 97900-000 RS, Brazil
}

\date{Accepted XXX. Received YYY; in original form ZZZ}

\pubyear{2022}

\begin{document}
\label{firstpage}
\pagerange{\pageref{firstpage}--\pageref{lastpage}}
\maketitle

\begin{abstract}
We present a detailed analysis of the ionised gas distribution and kinematics in the inner $\sim$ 200 pc of NGC\,4546, host of a Low Luminosity Active Galactic Nucleus (LLAGN). Using GMOS$-$IFU observations, with a spectral coverage of 4736$ - $6806 \AA \, and an angular resolution of 0.7 arcsec, we confirm that the nuclear emission is consistent with photoionisation by an AGN, while the  gas in the circumnuclear region may be ionised by hot low-mass evolved stars. The gas kinematics in the central region of NGC\,4546 presents three components: (i) a disc with major axis oriented along a position angle of $43^\circ \pm 3^\circ$, counter rotating relative to the stellar disc; (ii) non-circular motions, evidenced by residual velocities of up to 60 km s$^{-1}$, likely associated to a previous capture of a dwarf satellite by NGC\,4546; and (iii) nuclear outflows in ionised gas, identified as a broad component ($\sigma \sim 320$ km s$^{-1}$) in the line profiles, with a mass outflow rate of $\dot{M}_{\rm out} = 0.3 \pm 0.1$ M$_\odot$ yr$^{−1}$ and a total mass of $M_{\rm out} = (9.2 \pm 0.8) \times 10^3$ M$_\odot$ in ionised gas, corresponding to less than 3 per cent of the total mass of ionised gas in the inner 200 pc of NGC\,4546. The kinetic efficiency of the outflow is roughly 0.1 per cent, which is smaller than the outflow coupling efficiencies predicted by theoretical studies to AGN feedback become efficient in suppressing star formation in the host galaxy.

\end{abstract}

\begin{keywords}
Galaxies: active -- Galaxies: kinematics and dynamics -- Galaxies: individual (NGC\,4546)
\end{keywords}



\section{Introduction}
\label{sec:introduction}

It is well known that all galaxies with a spheroidal component (e.g. elliptical, lenticular and the bulge of spiral galaxies) host a supermassive black hole (SMBH) at their centre \citep{magorrian1998}. 
At some point in their life, SMBHs become active, giving rise to an Active Galactic Nucleus (AGN). Some AGN release huge amounts of energy from the SMBH accretion disc (e.g. quasars), placing them among the most luminous objects in the Universe. 

The most common type of AGN found in nearby early-type galaxies (ETGs) is the Low-Ionisation Nuclear-Emission Regions \citep[LINERs;][]{heckmann1980,1983Ferland_Netzer,1983Halpern_Steiner,Ho_Filippenko1997_V}, which is the lowest luminosity class among the AGN. Their optical spectra are dominated by narrow and strong emission lines of low ionisation species with line widths similar to those observed in the narrow-line region of Seyfert galaxies \citep[e.g.][]{heckmann1980,veilleux_osterbrock_1987}. However, LINER-type emission can also be produced by other excitation mechanisms, such as ionisation by shock waves \citep{heckmann1980} or by a population of hot low mass evolved stars \citep[HOLMES;][]{Binette_1994,Stasinska_2008}.

One way to confirm nuclear activity in LINERs is the detection of radio or X-ray cores in the host galaxy \citep{Ho_2008}. The presence of a broad component on permitted emission lines, such as H$\alpha$, is also a clear signature of an AGN (LLAGN, \citealt{Ho_Filippenko_1997_IV}). Nevertheless, given the weakness of the central engine, the observation of all these features is not straightforward. High spatial resolution data are usually necessary to isolate the nuclear emission from circumnuclear one in both X-rays and radio \citep{1989MNRAS.240..591Sadler,1994MNRAS.269..928Slee,Ho_2008, 2017ApJ...835..223She}. For the detection of the broad component in LLAGN, a proper subtraction of the stellar component is necessary \citep{Ho_Filippenko_1997_IV, Ricci2014b}, as in most cases the broad component emission is very faint.

Some studies reveal that ETGs have LINER-like ionised gas emission, which is often extended and can be observed up to kpc scales \citep{1986AJ.....91.1062Phillips, 1989ApJ...346..653Kim, sarzi2006,sarzi_2010, 2012ApJ...747...61Yan_Blanton}. 
In the case of these extended LINERs, the AGN as the dominant gas excitation mechanism over all the emission observed in ETGs can be discarded \citep{2012ApJ...747...61Yan_Blanton, 2013A&A...558A..43Singh,2016MNRAS.461.3111Belfiore}. 
This is due to the fact that the central AGN in LINERs is very weak, and it is not able to reproduce the intensity of the observed optical emission lines in the most extended regions. 
But, it does not mean that objects with LINER emission do not have an AGN at their centre, just that it is not powerful enough to ionise the total observed gas. So far, HOLMES are the best candidates to explain extended emission in LINERs \citep{2012ApJ...747...61Yan_Blanton, 2013A&A...558A..43Singh,sarzi_2010,Ricci2015_III,2016MNRAS.461.3111Belfiore}. But when it comes to nuclear LINERs, that is not so clear anymore.

A way to investigate the origin of the gas emission in LINERs is by using integral field spectroscopy (IFS) of the central region of LINER hosts. Besides allowing the study of the gas excitation mechanisms, such observations can be used to map the gas kinematics and investigate the effects that the AGN may have on the host galaxy.
The feedback is the effect of the AGN on the host galaxy due to the energy released (radiation and mechanical)  during the process of accretion of matter by the SMBH and which interacts with the host galaxy interstellar medium \citep[ISM;][]{fabian2012}. 
Signatures of feedback usually manifest in the form of winds (radiative mode) or in the form of powerful jets of particles observed in the radio (mechanical mode).
The radiative mode is mainly observed in galaxies with younger stellar populations, which host the most luminous AGNs at any epoch \citep[e.g. young quasars;][]{fabian2012,2017NatAs...1E.165Harrison}, while the mechanical mode is often seen in high-density environments, like on massive elliptical galaxies at the centre of galaxy clusters in the local Universe. 

AGN feedback is expected to play a major role in regulating the star formation rate.  In both radiative and mechanical feedback modes, outflows and jets are responsible for regulate star formation rates in the host galaxy in different scales by quenching or triggering the formation of new stars \citep{fabian2012}. In some cases, star formation can be triggered by compressed gas due to AGN winds \citep[positive feedback;][]{2013MNRAS.433.3079Zubovas,gallagher19} and more commonly, star formation can be suppressed by AGN winds and jets that maintain gas heated (maintenance mode) or remove the gas supply of the host galaxy \citep[negative feedback;][]{2009Cattaneo,fabian2012,2017NatAs...1E.165Harrison}. 
Due to its effects on star formation, AGN feedback is invoked to explain 
the formation and evolution of massive galaxies, since it may be responsible for the mass difference seen between models and observations \citep{2005Natur.433..604DiMatteo,fabian2012,Harrison_2018}. However, it is not clear how accretion onto the SMBH couples energy and momentum into the surrounding ISM to quench star formation, especially in low luminosity AGNs \citep{2013ARA&A..51..511Kormendy,Harrison_2018}.

In the brightest AGNs, it is possible to observe feedback effects at scales that can exceed the boundaries of the host galaxies. On the other side, in LLAGN, this phenomenon is more difficult to be observed, since the accretion rates in these cases are low, and, consequently, the energy released in the form of radiation and winds is much lower. 
This is specially true for LINERs, once AGN feedback can be seen only in the inner region of the host galaxies \citep[$\sim$ 100 pc;][]{Molina_2018}. 
Considering that LINERs are the most common type of AGN in the local Universe, a proper characterization of the feedback processes in these objects may help to expand our knowledge about AGNs and their effects on the host galaxies.

In this work, we aim to analyse the gas distribution and kinematics of the galaxy NGC\,4546, known to host a LLAGN based on optical diagnostic diagrams \citep{Ricci2014b}, in order to identify and characterize AGN driven winds in the inner region of this object. NGC\,4546 is a nearly edge-on lenticular galaxy that makes part of the Virgo V cluster and is located at about 17.8 Mpc of distance \citep{Galleta1987}. This object is characterized by the gas and stellar counter-rotation \citep{Galleta1987} in distinct planes \citep{Ricci2014a} and a possible external gas enrichment \citep{Galleta1987,Sage1994}, probably from a dwarf galaxy \citep{Sage1994,2020MNRAS.493.2253Escudero}.
NGC\,4546 presents a LINER-type nuclear activity, observed for the first time by \cite{sarzi2006} and confirmed through observations at 1.4 GHz \citep{sarzi_2010} and 5 GHz \citep{Nyland_2016} obtained with the Very Large Array (VLA), by unresolved nuclear radio emission, consistent with an accreting SMBH. 
\cite{Ricci2015_III} demonstrate that this galaxy has an ionised gas disc that reveals distortions in its radial velocity maps across the entire field of view, consistent with observations by \cite{sarzi2006} who detected similar distortions in the gas velocity fields but on larger scales, of the order of 20 arcsec. These perturbations can be caused by non-Keplerian effects, such as outflows or inflows, as suggested by \cite{sarzi2006} and \cite{Ricci2015_III}, and will be further explored in this work. 

This paper is organized as follows. In Section 2, we describe the observation, data reduction and the procedures applied to the final data. 
In Section 3, we present the results for the flux distribution and kinematics of the gas and stars. In Section 4, we discuss the results and, in Section 5, we present the conclusions of this work.

\section{Data and Measurements} \label{sec:observation}

\subsection{Observations and data reduction}
The data of NGC\,4546 were obtained on February 17th, 2008 (programme GS-2008A-Q-51) using the Integral Field Unit (IFU) coupled to the Gemini Multi-Object Spectrograph (GMOS) \citep{AllingtonSmith_2002} of the Gemini South Telescope. This galaxy was observed in the one-slit mode, which resulted in a data cube with a field-of-view (FOV) of 3.5$\times$5 arcsec$^2$. The R831-G5322 grating was used, covering a spectral range of 4736-6806 \AA, centred at 5800 \AA \, and with a spectral resolution of 1.3 \AA, estimated by measuring the full width at half maximum (FWHM) of the O\,{\sc i} $\lambda$5577 sky line \citep{Ricci2014a}. The seeing of the observation is 0.7 arcsec, estimated from the FWHM of field stars in the acquisition images taken through the $r$ filter.

The data reduction was performed using the standard {\sc iraf} package for the Gemini telescope with the following steps: bias and flat-field corrections, removal of cosmic rays with the LACOS algorithm \citep{van_Dokkum_2001}, wavelength and flux calibration and, building the data cube with a pixel scale of 0.05 arcsec \citep{Ricci2014a}. After the standard data reduction procedure, high- and low-frequency noise from both spatial and spectral dimensions were removed from the data cube of NGC\,4546 using the techniques described in \citet{menezes2019} (but see also \citealt{Ricci2014a}). 
Details on the data reduction and treatment can be found in \cite{Ricci2014a} and \citet{menezes2019}.

\subsection{Measurements} 
\label{sec:measurements}

For a detailed analysis of the gas distribution and kinematics, we use the Penalized Pixel-Fitting ({\sc ppxf}) method \citep{Cappellari2017} to subtract the stellar component from the observed spectra. 
To do this, we fitted the stellar absorption spectra assuming a line-of-sight velocity distribution described by the Gauss-Hermite series using as stellar templates those from the MILES-HC library \citep{Westfall_2019}. This library was created by applying the Hierarchical Clustering technique to the 985 MILES stars \citep{2006MNRAS.371..703SanchezBlazquez,2011A&A...532A..95FalconBarroso} and has a similar spectral resolution of our GMOS data of NGC\,4546. From this procedure, we obtained measurements of stellar kinematic parameters, such as maps of radial velocity, velocity dispersion and the Gauss-Hermite moments $h_3$ and $h_4$.

To obtain information about the gas distribution and kinematics, we used the {\sc ifscube} package, an analysis tool for Integral Field Spectroscopy data \citep{daniel_ruschel_dutra_2020_3945237,ruschel-dutra21}. This package allows the fit of the emission-line profiles by Gaussian curves or Gauss-Hermite series. 
To perform the fit, it is necessary to provide a configuration file containing the initial guesses (e.g. centroid velocity and velocity dispersion) for the emission lines that will be fitted.
Then, the fit starts with the spaxel corresponding to the centre of mass and proceeds in a spiral loop.
We fitted the emission lines using a residual data cube obtained by subtracting the contribution of the stellar component from the observed spectra. 
With the \textit{refit} option, the parameters of a previous successful fit within  a defined radius -- chosen to be 0.2 arcsec -- from the spaxel being fitted are to be used as initial guesses.
We set the emission lines ratio of [O {\sc iii}]\,$\lambda$5007\,/\,$\lambda$4959 and [N {\sc ii}]\,$\lambda$6548\,/\,$\lambda$6583 to their theoretical value 2.98 and 3.06 \citep{Osterbrock2006}, respectively, and tied the kinematics (velocity and velocity dispersion) of emission lines produced from the same species (H$\beta$ and H$\alpha$; [O {\sc iii}] $\lambda$5007 and [O {\sc iii}] $\lambda$4959; [N {\sc ii}]$\lambda$6548 and [N {\sc ii}]$\lambda$6583; [S {\sc ii}]$\lambda$6717 and [S {\sc ii}]$\lambda$6731).

From visual inspection of the data cube of  NGC\,4546, it is clear that the observed emission-line profiles present deviations from a single Gaussian curve at some locations. Thus, first, we fitted each emission line by Gauss-Hermite series and by two-Gaussian curves. 
Gauss-Hermite series are able to describe small deviations from a Gaussian profile, as wings and asymmetries. The observed line profiles in NGC\,4546 are well reproduced by the Gauss-Hermite series in all locations of the field of view. This procedure provides information on the average physical properties of the emitting gas. 
On the other hand, by fitting the emission lines by two-Gaussian curves, distinct kinematic components can be studied separately. Two components are clearly present in the observed profiles in the inner 1 arcsec radius, as  broad and  narrow components.  In regions farther from the nucleus, the line profiles present a single narrow component. At locations where the amplitude of the broad components is smaller than 3 times the standard deviation of the continuum noise next to the considered emission line, we fitted the profile by a single component. Spaxels where the amplitude of this component is not above 3 sigmas of the noise level are masked out from the maps.

\begin{figure*}
    \centering
    \includegraphics[width=0.85\textwidth]{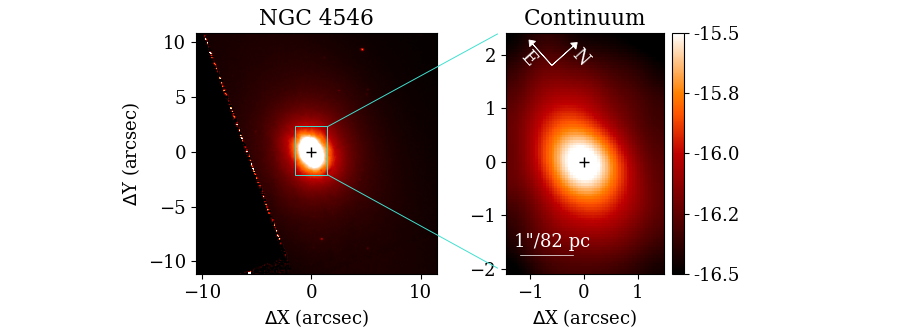}
    \includegraphics[width=\textwidth]{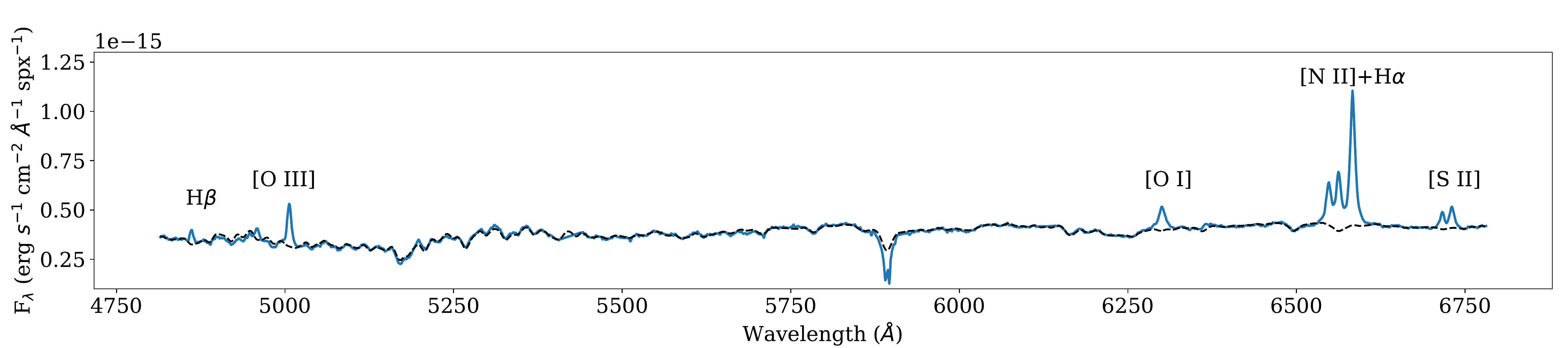}
    \includegraphics[width=0.49\textwidth]{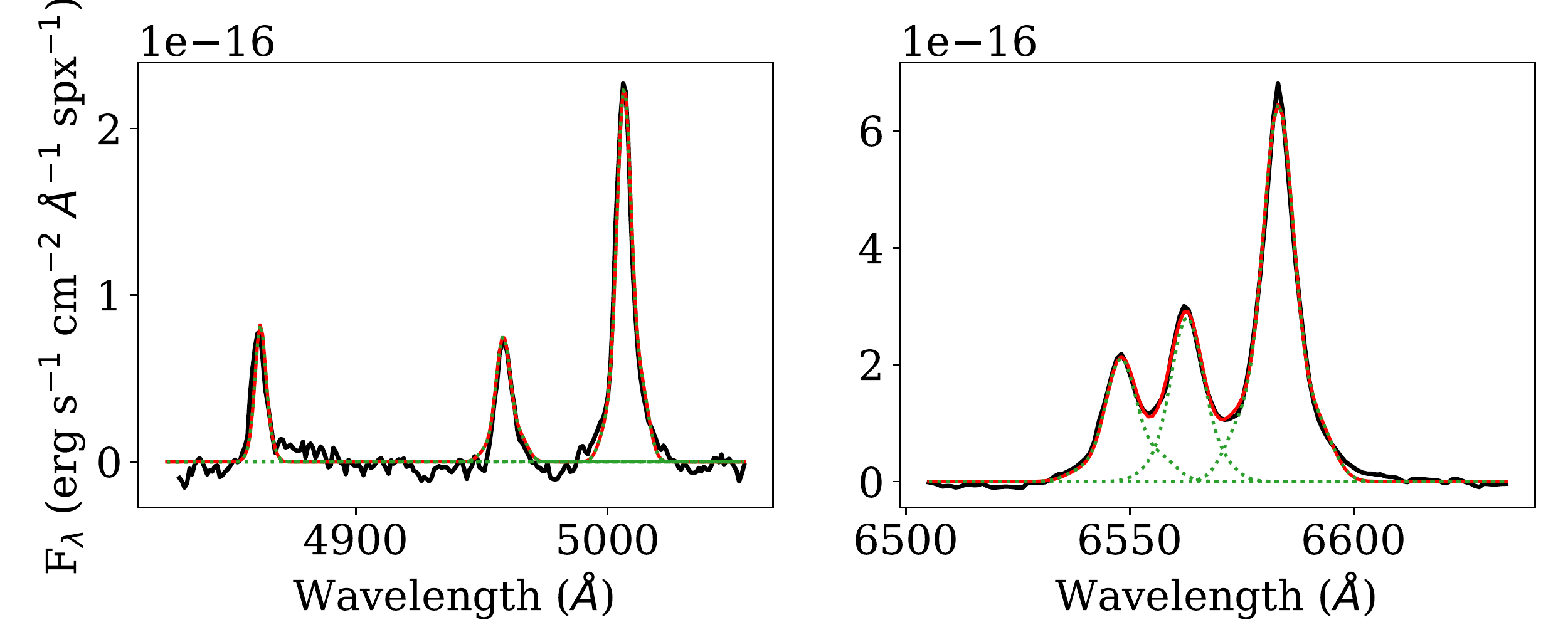}
    \includegraphics[width=0.49\textwidth]{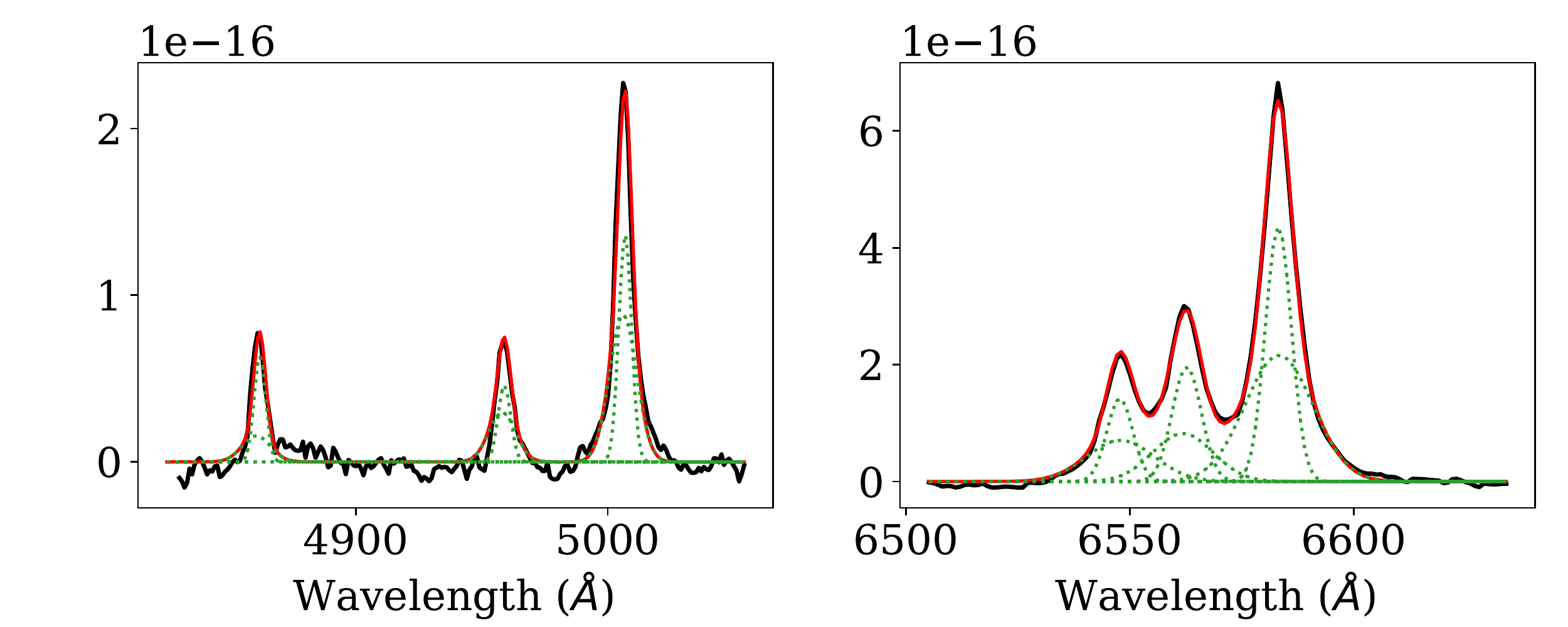}
    \caption{Top left: Image from NGC 4546 galaxy in large scale from the Hubble Space Telescope (HST) with the WFPC2-F606W instrument in the V band, obtained in the Hubble Legacy Archive. The blue rectangle indicates the FOV of the GMOS-IFU observation, with spatial dimensions of 3.5 arcsec x 5 arcsec. Top right: Continuum map of NGC 4546 in erg cm$^{-2}$ s$^{-1}$ \AA$^{-1}$ spaxel$^{-1}$ and in logarithmic scale. The continuum peak is represented by the black cross.
    Middle: Observed spectrum at the spaxel corresponding to the continuum peak with flux in erg s$^{-1}$ cm$^{-2}$ \AA $^{-1}$ spaxel$^{-1}$. The blue line represents the observed spectra and the stellar model spectra is represented by the black dotted line that was fitted using the {\sc ppxf} method \citep{Cappellari2017}. 
    Bottom: Examples of fits of the emission-line profiles for the nuclear spaxel, by Gauss-Hermite series (first two panels) and by two-Gaussian curves (last two panels). The black line shows the observed spectra, after subtracting the stellar component contribution. The red line shows the best fit models. For the Gaussian-fits, the individual components are shown as dotted green lines. }
    \label{fig:figure1}
\end{figure*}

The top left panel of Figure \ref{fig:figure1} shows a large-scale image of the NGC\,4546 galaxy in the V band, obtained with the Hubble Space Telescope (HST) using the  WFPC2-F606W instrument, taken from the Hubble Legacy Archive (HLA). The blue rectangle represents the observation field of the GMOS instrument. The top right panel shows the continuum map obtained from the GMOS-IFU data cube, performing an average of the continuum emission fitted by the {\sc ppxf} code for each spaxel. We define the position (0,0) in all maps as the continuum peak represented by the black cross. It is observed that the continuum emission in the central region of the galaxy follows the morphology on a large scale.
The middle panel of Figure \ref{fig:figure1} shows the observed nuclear spectrum in blue and the black dotted line represents the stellar component fitted with {\sc ppxf}. The main observed emission lines are identified: [O\,{\sc iii}]\,$\lambda\lambda4959,5007$, H\,$\beta$, [O\,{\sc i}]\,$\lambda6300$, [N\,{\sc ii}]\,$\lambda\lambda6548,6583$, H$\alpha$ e [S\,{\sc ii}]\,$\lambda\lambda6717,6731$. These lines are detected in almost the entire GMOS FOV, which enable us to fit the emission lines profiles and obtain point-to-point measurements of flux, velocity and velocity dispersion in the galaxy. Examples of fits of the line profiles are shown in the bottom panels for the nuclear spaxel.

\section{Results} \label{sec:results}

In Figure \ref{fig:maps_gh} we present the two-dimensional maps for the stellar and gas flux distributions and kinematics. The stellar kinematic maps are based on the {\sc ppxf} fits, while the emission-line properties are obtained by fitting the observed line profiles by Gauss-Hermite series, as described previously. We show the flux distributions and kinematic maps for the [O\,{\sc iii}]\,$\lambda$5007, [O\,{\sc i}]\,$\lambda$6300, H$\alpha$\,$\lambda$6563, [N\,{\sc ii}]\,$\lambda$6583 and [S\,{\sc ii}]\,$\lambda$6716 emission lines, which are the strongest lines among their species.  Grey regions correspond to masked out locations, where the amplitude of the corresponding emission line is smaller than 3 times the standard deviation of the continuum in regions close to the line. We subtracted the systemic velocity of the galaxy of $1046$ km s$^{-1}$ from the observed stellar and gas velocity fields, as obtained by modelling the stellar velocity field with the {\sc kinemetry} method \citep{Krajnovic2006_kinemetry}.
We also corrected the velocity dispersion values by the instrumental broadening. 
Below we describe the maps for the stellar kinematics and, mainly, the results for the emission line properties obtained by fitting the line profiles  by Gauss-Hermite series and two Gaussian components.

\subsection{Stellar kinematics}

The top row of Fig.\,\ref{fig:maps_gh} shows the maps for the stellar kinematics of NGC\,4546. 
The stellar continuum emission was obtained by performing a flux average in a spectral window from 4815 \AA\, to 6780 \AA\, per spaxel. The stellar velocity field exhibits a well-behaved rotation pattern, with amplitudes in the order of $100\pm15$ km s$^{-1}$ and the line of nodes orientated in the east-west direction. The stellar velocity dispersion ($\sigma_\star$) map shows values in the range of 160 to 260 km\,s$^{-1}$, with the highest values seen at the nucleus and along the north-south direction (the minor axis of the galaxy). The $h_3$ map shows negative values at locations where the redshifts (relative to the systemic velocity of the galaxy) are observed in the velocity field and positive values in blueshifted regions. This anti-correlation may be produced by the movement of the bulge stars lagging the rotation as found in previous works \citep{emsellem06,Ricci2016,rogemar17_stellar}. The $h_4$ map shows values smaller than 0.05 in most locations. Maps for the stellar velocity, velocity dispersion and $h_3$ moment for NGC\,4546, based on the same data used here, were already presented in \citet{Ricci2016} as part of a study aimed to discuss general properties of the circumnuclear stellar kinematics in 10 early-type galaxies. The maps presented here are consistent with those shown in \citet{Ricci2016}. Therefore, we only present the maps for the stellar kinematics to allow a direct comparison with the gas kinematics and we will not analyze the stellar kinematics in details.

\subsection{Gas emission}

\begin{figure*}
    \centering
    \includegraphics[width=\linewidth]{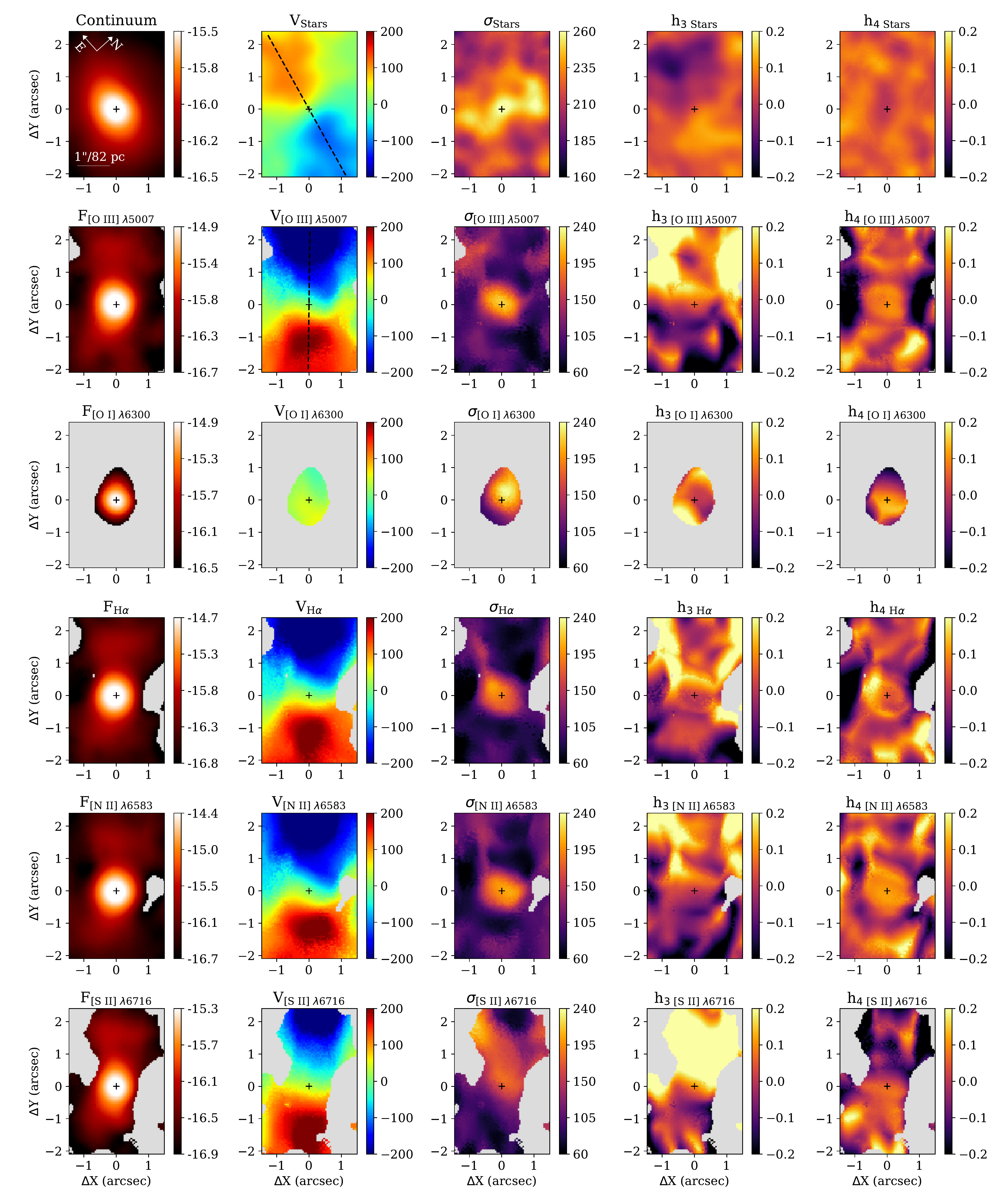}
    \caption{Continuum, flux distribution and kinematics two-dimensional maps of the brightest optical emission lines and the stellar component. First column: Continuum and flux distributions in logarithm scale. Second column: observed velocity field in km s$^{-1}$. Third column: dispersion velocity in km s$^{-1}$. Fourth and fifth: the Gauss-Hermite moments. The continuum peak is represented by the black cross. The grey colours on the maps are the regions that we mask to hide the bad measurements by applying a cut in amplitude greater than three times the standard deviation of the stellar continuum next to the line. The dashed black line indicates the orientation of the major axis of the gas and stellar discs.}
    \label{fig:maps_gh}
\end{figure*}

The emission-line flux maps are shown in the first column of Figure \ref{fig:maps_gh}, in logarithmic scale. 
The [O\,{\sc iii}], H$\alpha$, [N\,{\sc ii}] and [S\,{\sc ii}] maps present extended emission over most of the GMOS field of view, with the highest flux levels observed along the northeast--southwest direction, which is displaced from the orientation of the major axis of the galaxy by $30^\circ \pm 6^\circ$ as obtained by comparing the emission-line flux distributions with the continuum map (top-left panel of Fig.~\ref{fig:maps_gh}). The [O\,{\sc i}]\,$\lambda$6300 emission-line is detected above 3$\sigma$ of the continuum noise only in the inner 1 arcsec.  The peak of emission of all emission lines are observed at the nucleus of the galaxy. 

\begin{figure*}
    \centering
    \includegraphics[scale=0.45, trim={0 0 6.5cm 0}, clip]{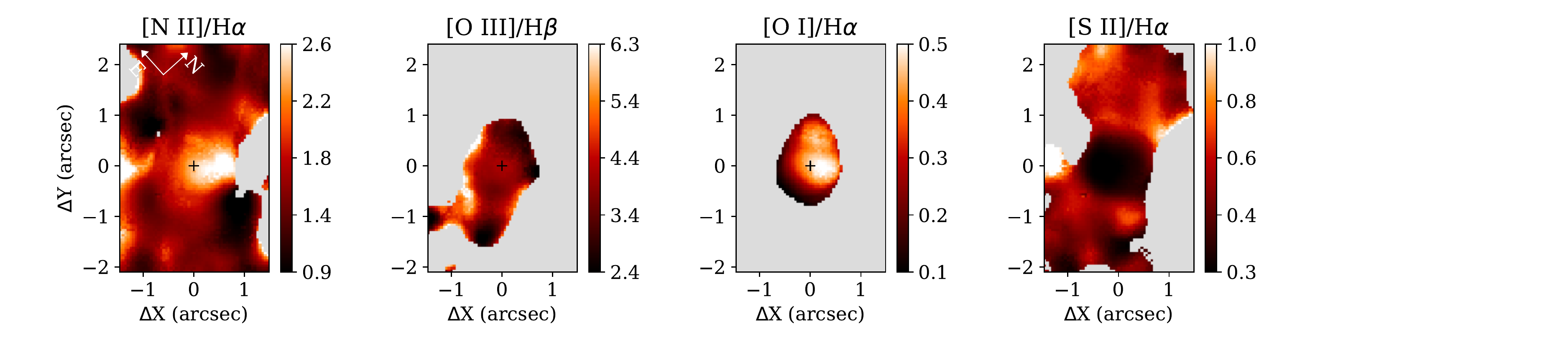}
    \caption{Emission line ratios maps for the galaxy NGC 4546. First panel: [N\,{\sc ii}]\,$\lambda$6583/H$\alpha$ line ratio. Second panel: [O\,{\sc iii}]\,$\lambda$5007/H$\beta$ line ratio. Third panel: [O\,{\sc i}]\,$\lambda$6300/H$\alpha$ line ratio. Fourth panel: [S\,{\sc ii}]\,$\lambda$6716/H$\alpha$ line ratio.}
    \label{fig:line_ratio}
\end{figure*}

In Figure \ref{fig:line_ratio} we present the [N\,{\sc ii}]/H$\alpha$, [O\,{\sc iii}]/H$\beta$, [O\,{\sc i}]/H$\alpha$ and [S\,{\sc ii}]/H$\alpha$ flux ratio maps. These line ratios are useful to map and investigate the gas excitation in the inner region of NGC\,4546. The [N\,{\sc ii}]/H$\alpha$ map, shown in the first panel, is similar to the one presented by \citet{Ricci2015_III}, based on the fitting of the emission-lines by Gaussian curves. Our [N\,{\sc ii}]/H$\alpha$ map shows the highest values at $\approx$ 0.5 arcsec northwest of the nucleus, while \citet{Ricci2015_III} found the highest ones in the nucleus.  This slight difference is likely due to the different methods used to fit the emission line profiles, which are not well reproduced by a single Gaussian in the central region. 
In the second panel, we present the [O\,{\sc iii}]/H$\beta$ ratio map, in which we can observe values $\geq$ 2.5 over the entire field. 
Although the  [O\,{\sc i}]\,$\lambda$6300  emission is detected only in the central region, the [O\,{\sc i}]/H$\alpha$ ratio map (third panel) shows the highest values, of up to 0.5,  at $\approx$ 0.5 arcsec northwest of the nucleus, coincident with the highest [N\,{\sc ii}]/H$\alpha$ structure. The [S\,{\sc ii}]/H$\alpha$ map presents the smallest values in the inner 0.5 arcsec ([S\,{\sc ii}]/H$\alpha\approx$ 0.3), while the highest values of up to 1.0 are observed in locations further away from the nucleus.

\begin{figure}
    \centering
    \includegraphics[width=\linewidth]{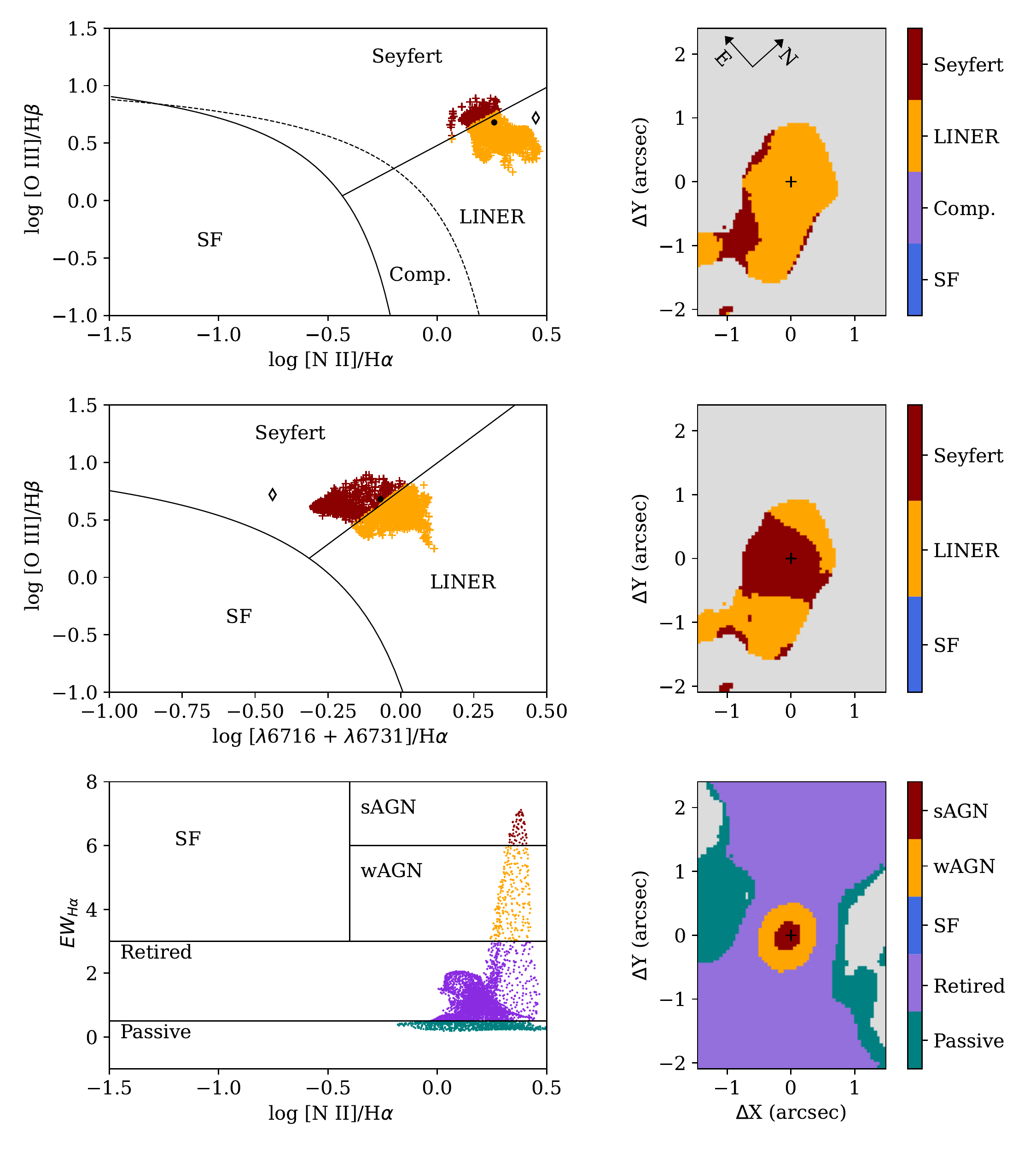}
    \caption{
    Diagnostic diagrams and excitation maps for the NGC\,4546 galaxy. Top panel: BPT diagnostic diagram \citep{bpt_1981} for the [N\,{\sc ii}]/H$\alpha$ emission line ratio. The continuous black curve was proposed by \citet{kauffmann_2003} and separates SF and transition objects, the black dotted curve separates transition objects and AGN and was proposed by \citet{kewley2001b} and the black line is the division between LINER and Seyfert emission from \citet{Cid_Fernandes_2010}. Middle panel: BPT diagnostic diagram for the [S\,{\sc ii}]/H$\alpha$ emission line ratio. 
    The black curve divides SF and AGN emission \citep{kewley2001b} and the black line separates LINER from Seyfert emission \citep{Kewley_2006}. 
    The black dots represent line ratios for the disc component. For each emission line, the contribution of the disc was obtained as the subtraction of the fluxes of the broad component measured in the nuclear spectrum (integrated within a 0.7 arcsec aperture) from the fluxes measured using the integrated spectrum over the entire FOV. The black open diamonds represent the line ratios obtained for the outflow (the broad component). 
    Bottom panel: WHAN diagram for H$\alpha$ equivalent width vs. [N\,{\sc ii}]/H$\alpha$ line ratio \citep{whan_cid_fernandes_2011}.
    }
    \label{fig:diagnostic}
\end{figure}

The gas excitation mechanisms can be investigated using the BPT \citep{bpt_1981}  and  WHAN \citep{whan_cid_fernandes_2011} diagnostic diagrams. We present the  [O\,{\sc iii}]/H$\beta$ vs. [N\,{\sc ii}]/H$\alpha$ and  [O\,{\sc iii}]/H$\beta$ vs. [S\,{\sc ii}]/H$\alpha$ diagrams in the top-left and middle-left panels of Figure \ref{fig:diagnostic}, respectively, and the WHAN diagram in the bottom panel of Figure \ref{fig:diagnostic}. We do not present the [O\,{\sc i}]-based BPT diagram because the [O\,{\sc i}]\,$\lambda$6300 emission is detected only in the very central region of the galaxy. In the right panels, the colour coded excitation maps are shown according to the location of each spaxel in the diagnostic diagrams. The [N\,{\sc ii}]-based BPT diagram indicates that the whole field of view has an emission typical of LINERs. For the [S\,{\sc ii}]-based BPT diagram, one may see that the nuclear region of the galaxy shows line ratios typical of Seyfert-like AGNs. Note that the WHAN diagram also indicates that the nucleus of NGC\,4546 is associated with an AGN. This result is consistent with previous optical \citep{sarzi2006,Ricci2014b} and radio observations  \citep{sarzi_2010,Nyland_2016} of this galaxy. At larger distances from the nucleus, the H$\alpha$ equivalent width (EW$_{H \alpha}$) is smaller than 3 indicating an ionization mecanism similar to that of 'retired' galaxies \citep{whan_cid_fernandes_2011}. Retired galaxies are populated mainly by old stars, with very low star-formation, and their nebular emission is usually classified as LINERs. However, the ionizing source of these objects is related to HOLMES rather than nuclear activity \citep{Binette_1994,Stasinska_2008,whan_cid_fernandes_2011}.

\subsection{Gas kinematics}

The second column of Figure \ref{fig:maps_gh} presents the stellar and gas velocity fields. 
It is possible to observe that the rotation of the stellar component and the gas have nearly opposite directions (displaced by $\sim135^\circ$), as previously seen in other works \citep{Galleta1987, Ricci2014a}. The gas velocity fields for all emission lines are similar and present blueshifts to the northeast of the nucleus and redshifts are observed southwest of it. The gas velocity amplitude is $210\pm25$ km\,s$^{-1}$, twice the velocity amplitude of the stars. 
In addition, while the stellar velocity field shows a well-behaved rotation pattern and is in agreement with previous works \citep{Ricci2016, Ricci2020}, the gas clearly presents deviation of a pure disc rotation pattern. 
The gas velocity dispersion maps (third column of Figure \ref{fig:maps_gh}) present values in the range 60--240 km\,s$^{-1}$,  smaller than the stellar velocity dispersion at most locations. The highest velocity dispersion values for the gas (of up to 240 km\,s$^{-1}$) are seen in the inner $\sim$1 arcsec.

The Gauss-Hermite moments $h_3$ and $h_4$ measure asymmetric and symmetric deviations of a Gaussian line profile \citep{1993van_der_Marel_Franx, 1993Gerhard,Riffel2010}, respectively.
The $h_3$ moment maps show absolute values of up to 0.2 for all emission lines, but unlike what is observed for stars, there is no anti-correlation between $h_3$ and the velocity field for the gas. Positive $h_3$ values are observed mainly to the northeast of the nucleus, while negative values are more common in the southwest side of the galaxy. 
The $h_4$ moment is positive over most of the field of view with values of up to 0.2, while some negative values are also observed at some locations. The high absolute values of $h_3$ and $h_4$ observed for all emission lines indicate that the line profiles present a significant deviation of a pure Gaussian curve.

\subsection{2D maps from two-Gaussian fits of the line profiles}

\begin{figure}
    \centering
    \includegraphics[width=\linewidth]{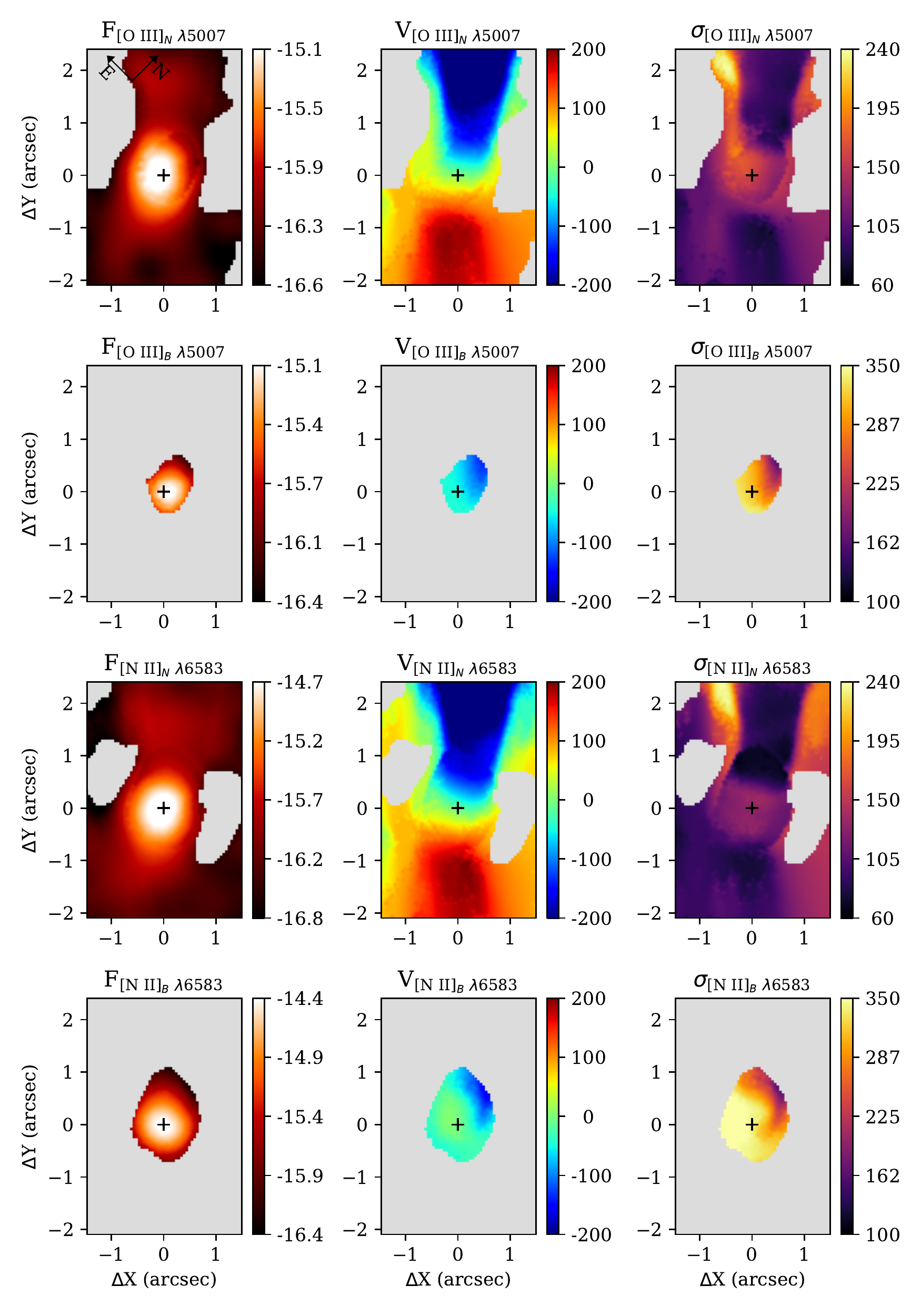}
    \caption{Flux and kinematics maps resulting from the Gaussian fit. First and second panels: maps for the narrow and broad components of the [O\,{\sc iii}]\,$\lambda$5007 emission line, respectively. Third and fourth panels: maps for the narrow and broad components of the [N\,{\sc ii}]\,$\lambda$6583 emission line, respectively. The grey colours on the maps are regions that we mask to hide the bad measurements by applying a cut in amplitude greater than three times the standard deviation of the continuum next to the emission line. }
    \label{fig:maps_2g}
\end{figure}

As mentioned above, the maps of $h_3$ and $h_4$ for the emission lines show values that deviate significantly from zero, indicating that the line profiles present more than a single kinematic component. A visual inspection of the spectra shows that at least two kinematic components are present in the nuclear region, as can be seen in Fig. \ref{fig:figure1}. 
The fitting of the emission-line profiles by multi-Gaussian functions can be used to separate different kinematics components that  account for the observed emission. For instance, \citet{2021A&A...645A.130Comeron} used up to six different components to reproduce the emission-line profiles  in the central region of the nearby Seyfert galaxy NGC\,7130, using high quality VLT-MUSE data. This indicates that the shape of emission line profiles produced by AGN can be much more complicated and advanced fitting techniques may be useful to properly extract the information on the gas emission structures, with high quality data from the next generation of telescopes. 
Thus, we fitted the emission lines with two Gaussian components in order to separate these two components following the procedure described Sec. \ref{sec:measurements}. The resulting two-dimensional maps for the  [O\,{\sc iii}]\,$\lambda$5007 and [N\,{\sc ii}]\,$\lambda$6583 emission lines are presented in Figure \ref{fig:maps_2g}. We separate the maps according to the widths of the Gaussians, which we will refer to as the narrow component and the broad component. It is worth mentioning here that what we call a broad component is unrelated to the AGN broad line region. Grey regions in each map correspond to locations where the corresponding component is not detected above 3$\sigma$ of the noise level. In regions where only the broad component is not detected, we fit the line profile by a single Gaussian function, corresponding to the narrow component.

The broad component is detected only in the inner 0.5 arcsec in [O\,{\sc iii}]\,$\lambda$5007  and in the inner 1 arcsec in [N\,{\sc ii}]\,$\lambda$6585. It is slightly blueshifted relative to the galaxy rest frame by 50\,--\,100 km\,s$^{-1}$ and its velocity dispersion ranges from 200\,--\,350 km\,s$^{-1}$, about twice the values seen for the narrow component at the same locations.

The flux, velocity and velocity dispersion maps of the narrow component are similar to those based on the Gauss-Hermite fits (Fig.~\ref{fig:maps_gh}) in regions where only the narrow component is detected. In the region where both components are detected, the narrow component contributes to roughly 50 per cent of the total flux of each line, the velocity fields of the narrow component are less disturbed than the velocity fields obtained by fitting the lines by Gauss-Hermite series and the velocity dispersion values of the narrow component are slightly smaller than those based on Gauss-Hermite series. Although the  $h_3$ and $h_4$  maps (Fig.~\ref{fig:maps_gh})  show high absolute values in regions away from the nucleus, which could indicate the presence of a secondary component in the line profiles, the broad component is not detected above 3$\sigma$ of the noise level.

\section{Discussion} \label{sec:discussion}

NGC\,4546 was already studied using optical IFU data at kpc scales as part of the SAURON project \citep[e.g.][]{emsellem06_sauronIII,sarzi2006} and at scales of hundreds of parsecs as part of a sample of 10 early-type galaxies observed with Gemini GMOS-IFU \citep[e.g.][]{Ricci2014a,Ricci2014b,Ricci2015_III,Ricci2016}. These studies focused on the general properties of their samples rather than the details of the distribution, excitation and kinematics of the gas in the central region of the galaxies. Here, we use the same GMOS data used in the works above to analyse the ionised gas emission in the inner 200 pc of NGC\,4546, and in particular, to explore the origin of the perturbations in the radial velocity maps previously reported in both small and large spatial scales \citep{Ricci2015_III,sarzi2006}.

We find that the gas radial velocity fields based on Gauss-Hermite fits (Fig. \ref{fig:maps_gh}) cannot be explained only by circular motions in the plane of the galaxy disc.  This is consistent with what was found by \cite{Ricci2015_III}, despite using different methodologies to reproduce the emission line profiles. 
The gas kinematics maps presented by \cite{Ricci2015_III} was created from fitting one Gaussian per emission line and by keeping the kinematics of [N\,{\sc ii}], H$\alpha$ e [S\,{\sc ii}] coupled. In addition, they included a weak broad component to reproduce the H$\alpha$ nuclear profile. 
The origin of these kinematic perturbations is still unknown, and previous works suggested they are associated with AGN-driven ionised winds \citep{sarzi2006,Ricci2015_III}.

Although \cite{Bettoni1991} suggests the presence of a small bar in this galaxy which could produce non-circular motions, more recent photometric studies show that there is no evidence of a bar component \citep{2018ApJ.862.100Gao,2020MNRAS.493.2253Escudero}. 
In addition, the recent photometric analysis revealed several components that may be associated with a bulge, disc/lenses and halo \citep{2018ApJ.862.100Gao,2020MNRAS.493.2253Escudero}, as well as revealing an extensive asymmetric dust structure along the galaxy's semi-major axis that extends up to 3.5 kpc from the nucleus \citep{2020MNRAS.493.2253Escudero}.
The possibility of a recent interaction of NGC\,4546 with a neighbouring ultra-compact dwarf \citep[UCD; ][]{2011EAS....48..259Norris&Kannappan,2015MNRAS.451.3615Norris} may be responsible for creating multiple structural components \citep{2018A&A...617A.113Eliche_Moral,2020MNRAS.493.2253Escudero} that were observed by \cite{2018ApJ.862.100Gao} and \cite{2020MNRAS.493.2253Escudero}.  
And this scenario of a merger event is plausible, given the star formation history (SFH) of the UCD companion and the evidence of co-rotation with respect to the gas in NGC\,4546  \citep{2011MNRAS.414..739Noris&Kannappan,2020MNRAS.493.2253Escudero}.

Regarding the gas kinematics, as mentioned above, the gas velocity fields based on Gauss-Hermite fits of the line profiles show important distortions from a disc rotation pattern. In addition, in the nuclear region to reproduce the line observed line profiles, a secondary broad component is needed. Such broad components are commonly used as signatures of ionised gas outflows \citep[e.g.][]{greene12,emonts17,perrota19,rogemar20_n1275}. In the following sections, we discuss the origin of the gas emission and the distinct kinematic components.

\subsection{Gas excitation}

It is not new that NGC\,4546 hosts an Active Galactic Nucleus. \cite{sarzi_2010} and \cite{Nyland_2016} find unresolved nuclear emission in radio at frequencies 1.4 GHz and 5 GHz, respectively, associated with an AGN.
In addition to the unresolved radio core, a study of the nuclear region of the galaxy NGC\,4546, \cite{Ricci2014b} detected a broad component in H$\alpha$, interpreted as further evidence that nuclear activity is present in this object. In this work, we do not include a broad component as it is not necessary to properly reproduce the line H$\alpha$ nuclear profile, as shown in Fig.~\ref{fig:figure1}.  In addition, \citet{Ricci2014a}, based on emission-line ratio diagnostic diagrams, found that the nuclear emission of NGC\,4546 is consistent with gas ionisation by an AGN. The origin of the extended gas emission was not included in previous studies. 

Although the presence of nuclear activity in ETG galaxies is well established, it is not known whether this activity is responsible for the emission of extended gas in these objects. Some studies reveal that AGN alone does not have the ability to ionise the gas at kpc scales, and another mechanism prevails over the gas excitation in these regions \citep{2012ApJ...747...61Yan_Blanton,2013A&A...558A..43Singh,sarzi_2010,2016MNRAS.461.3111Belfiore}. But what is not clear yet is the possibility of the AGN ionizing the circumnuclear regions (scales of $\sim$ 100 pc) since for such a study, it is necessary to have access to the nuclear region with high spatial resolution data. 
In view of this, \cite{Ricci2015_III} suggests that, although the AGN is not responsible for ionizing the gas disc, it can photoionise the gas in the perpendicular region. And this is one of the cases in which NGC\,4546 fits, as \cite{Ricci2015_III} observes a gas that he defines as low-velocity and which may be associated with an ionisation cone, possibly related with the AGN.

\cite{Ricci2014b,Ricci2015_III} found that the emission lines ratio of the nuclear and circumnuclear regions are typical of LINERs, based on the analysis of BPT \citep{bpt_1981} diagnostic diagrams using flux measurements based on integrated spectra from these regions. Here, we present the spatially resolved BPT and WHAN \citep{whan_cid_fernandes_2011}, which confirm that the nuclear emission of NGC\,4546 is consistent with emission of gas photoionised by an AGN. 

LINER emission is also confirmed using EW$_{H\alpha}$ with the aid of the WHAN diagram (Fig. \ref{fig:diagnostic}). As noted by \cite{Ricci2014b}, NGC\,4546 is at the limit of the LINER and Seyfert classification. 
In regions further away from the nucleus, the BPT diagrams show mainly LINER-like emission, while WHAN diagram shows mainly typical values of ``Retired Galaxies'' from spaxels at distances larger than $\sim$0.5 arcsec from the nucleus.  This indicates that the line emission from the circumnuclear region of NGC\,4546 may be produced by gas ionised by hot low-mass evolved stars (HOLMES) as proposed by previous works of early-type galaxies \citep{2012ApJ...747...61Yan_Blanton,2013A&A...558A..43Singh,sarzi_2010,Ricci2015_III,2016MNRAS.461.3111Belfiore}.

Summarizing, we confirm that the nuclear gas emission in NGC\,4546 is produced by an AGN, while the simultaneous analysis of the BPT and WHAN diagrams indicate that the emission from regions farther than 0.5 arcsec from the nucleus may be produced by HOLMES.

\subsection{Gas kinematics}

The kinematic maps for the emission lines show a disc component, with a misalignment between the planes \citep[PA$_{\rm stellar} - PA_{\rm gas} \sim -129^\circ$; ][]{Ricci2014a} and a counter-rotation of the ionised gas and stars \citep[PA$_{\rm stellar} - PA_{\rm gas} \sim - 144^\circ$; ][]{Galleta1987,sarzi2006}. Besides the rotation pattern, the gas kinematics reveals also non-circular motions. In addition, the emission-line profiles from the central region of  NGC\,4546 present a broad component that traces the emission of ionised outflows produced by the central AGN.

In the next subsections, to characterize the perturbations caused by non-circular motions in the disc of NGC\,4546, we use the {\sc kinemetry} method \citep{Krajnovic2006_kinemetry} applied to the observed velocity fields.
{\sc kinemetry} assumes that it is possible to describe the observed velocity field from a set of ellipses with a cosine law, whereas the velocity field is described by a circular motion in a thin disc. It is also possible to apply this method to the higher-order moments of the line-of-sight velocity distribution (LOSVD), like velocity dispersion and Gauss-Hermite moments $h_3$ and $h_4$. 
{\sc kinemetry} does this by using the Fourier expansion technique, with a methodology similar to what is used in surface photometry \citep{Krajnovic2006_kinemetry}. 
The {\sc kinemetry} applied to the velocity field returns the modeled velocity field and the ellipse's parameters, such as position angle ($\Psi_0$) and ellipticity ($\epsilon$), and the kinemetric coefficients ($k_n$) as a function of radius.
Regarding the ionised outflow, we estimate the mass of ionised gas, the mass outflow rate and kinetic power.

\subsubsection{The disc component and non-circular motions} \label{section:kinemetry_gas}

\begin{figure*}
    \centering
    \includegraphics[scale=0.45]{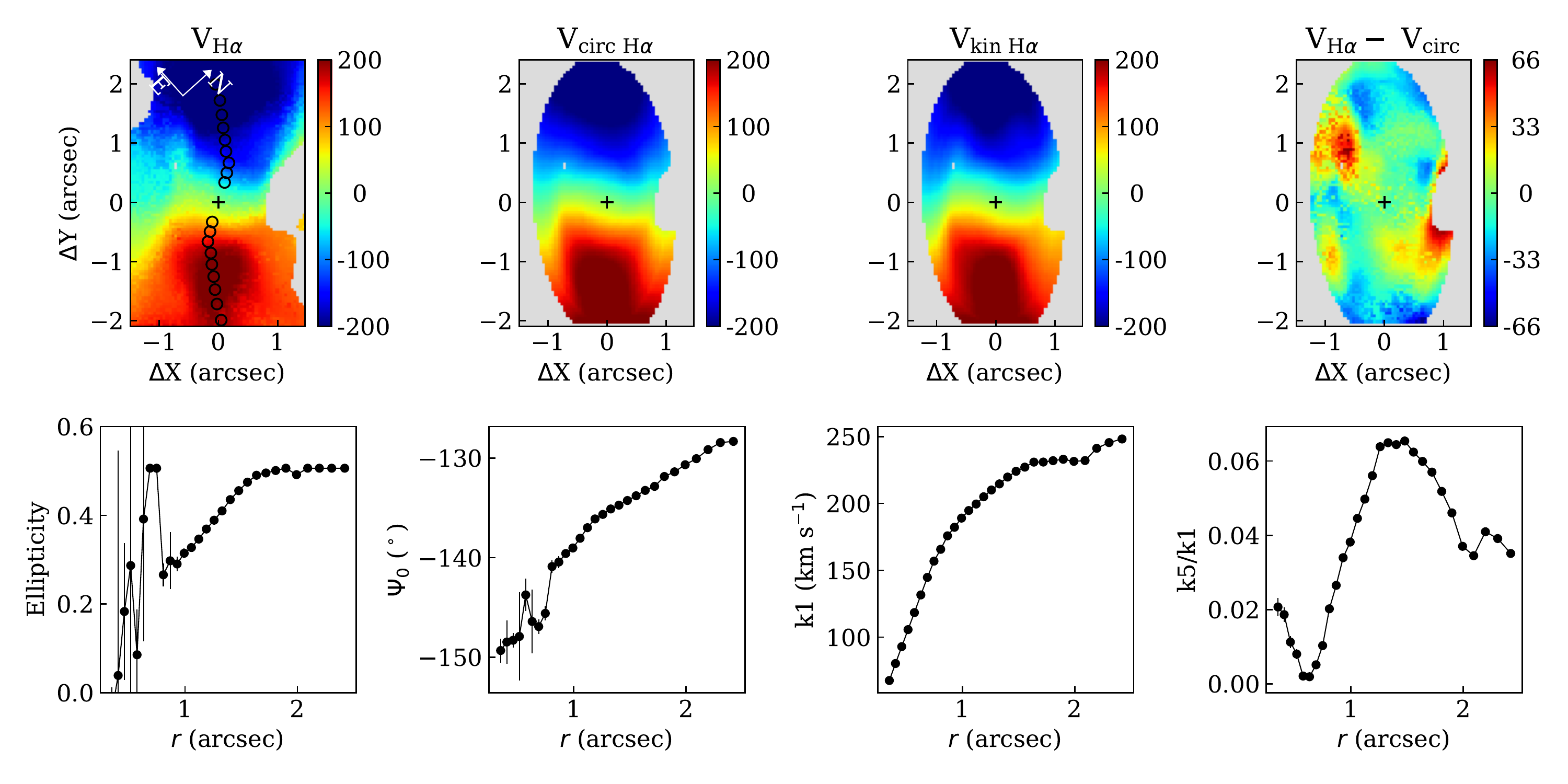}
    \caption{H$\alpha$ velocity maps and kinemetric coefficients resulting from the {\sc kinemetry} fit \citep{Krajnovic2006_kinemetry}. 
    From left to right in the top panel: observed gas velocity field, radial velocity, kinemetric velocity and residual velocity. The residual map is obtained by subtracting the observed velocity and the circular velocity fitted by the code. The black open circles indicate the position angle $\Psi_0$ of the modelled ellipses along the radius.
    From left to right in the bottom panel: ellipticity, semi-major axis position angle and the harmonic expansion coefficients $k_1$ and $k_5/k_1$.}
    \label{fig:kinemetry_gas}
\end{figure*}

In Figure \ref{fig:kinemetry_gas} we present the two-dimensional maps, ellipse parameters and kinemetric coefficients resulting from the fit using the {\sc kinemetry} method for the H$\alpha$ velocity field, obtained from the Gauss-Hermite fits of the line profile. 
From left to right in the top panel of Fig. \ref{fig:kinemetry_gas} are the observed velocity field, the circular velocity model, the kinemetric model and the residual map. And from left to right in the bottom panel are the radial profiles of ellipticity, position angle, k$_1$ and k$_5$/k$_1$ kinemetric coefficients. The coefficient $k_1$ is the dominant kinemetric term and describes the rotation curve, and the $k_5$ term describes motions that deviate from a rotation pattern and is generally normalized with respect to the $k_1$ term.

From the residual map, it is clear that this galaxy presents features that cannot be explained by circular motions since the circular velocity model (k$_1$ harmonic coefficient) is not enough to reproduce the observed map well. Higher-order terms (k$_3$ and k$_5$) of harmonic expansion are needed to reproduce it.
A large variation in the position angle of the line of nodes of the gas velocity field is observed with the radius ($\Delta \Psi_0 \sim 25^\circ$), with an average value of $\Psi_0=-137^\circ$. 

A large variation in the position angle of the line of nodes of the gas velocity field is observed with the radius ($\Delta \Psi_0 \sim 25^\circ$), with an average value of $\Psi_0=-137^\circ$. 
This value is in agreement with the value of $-149^\circ$ obtained by \citep{Ricci2014a} and $-137^\circ$ obtained by \citep{sarzi2006}. 
The ionised gas circular velocity has a mean ellipticity of 0.34 with a variation along the radius of $\Delta \epsilon \sim 0.4$ inside the inner 1.5 arcsec and remains constant until reaching the end of our FOV.

The radial velocity profile k$_1$, which describes the profile of a rotating disc, shows that the gas velocity increases until it reaches a maximum of $\sim$ 240 km s$^{-1}$ in the outermost region of the GMOS field of view. 
The $k_5/k_1$ ratio is maximum around 1.5 arcsec (130 pc) in order of 6 per cent and may be indicating a transition region between kinematic components \citep{emsellem06_sauronIII}.
The combination of behaviours revealed by the fitted ellipses across our FOV is that we have two components of ionised gas acting in the inner 2.5 arcsec (215 pc) of this galaxy. 
First, we have a central component that is mainly round ($\epsilon < 0.4$) with a change in the position angle of $\Delta \Psi \sim 15^\circ$. Secondly, is a component presenting nearly constant ellipticity ($\epsilon \sim 0.5$) and velocity, tracing the ionised gas disc. 
And the transition region, where $k_5/k_1$ reaches its maximum, corresponds to the region where twists are observed in the gas velocity field. This type of twists are commonly observed in cases where the galaxy has been enriched with external gas \citep{2017MNRAS.464.4227Raimundo,2021A&A...650A..34Raimundo}. What happens is that the interaction of the gas with the gravitational potential of the stars may cause loss of angular momentum, in which the gas in the innermost region aligns more quickly with the stars than the outermost region resulting in a disturbed disc \citep{2015MNRAS.451.3269VandeVoort}.

As we may recall, NGC\,4546 has a relatively close companion, a remnant galaxy that possibly has been through an interaction with the progenitor galaxy \citep{2011MNRAS.414..739Noris&Kannappan,2020MNRAS.493.2253Escudero}. 
The most plausible scenario for the observed high redshifted residuals ($\sim 60$ km s$^{-1}$), is that it may be the remnants traces of the interaction with the ultra-compact dwarf companion. 
This type of velocity residue can be associated with inflows of gas that are falling toward the nucleus \citep{2017MNRAS.470.1703Diniz}. Gas inflows have been previously observed in galaxies that have undergone some type of merger, usually a minor merger \citep{2015ApJ...799..234Fischer,2019MNRAS.486..691Brum}, where inflows can result from this type of event \citep{2015ApJ...799..234Fischer} and triggering nuclear activity in the major galaxy \citep{2014ApJ...792..101Davies,2015MNRAS.451.3587Riffel,2017MNRAS.464.4227Raimundo,2021A&A...650A..34Raimundo}.

\subsubsection{Ionised Gas Outflows} \label{section:outflows}

We note the need for a second component (a broader one) to describe the emission lines profiles in the nuclear region of the galaxy NGC\,4546. This component may be associated with an ionised outflow produced by the central AGN. In this sense, it is possible to estimate some properties of this outflow, such as its mass of ionised gas, its mass rate and kinetic power. These properties are important to understand how the outflow may affect the host galaxy, especially in low luminosity AGNs.

To derive the outflow properties, we use a nuclear spectrum integrated within a diameter of the seeing (0.7 arcsec) -- as the outflow is restricted to the nuclear region -- using the datacube free of the stellar continuum contribution. We fit the emission line profiles profiles by two Gaussian profiles using the {\sc ifscube} code \citep{daniel_ruschel_dutra_2020_3945237,ruschel-dutra21}, as described above. We also obtain an integrated spectrum within the whole GMOS FoV to obtain the properties of the disc and fit the emission lines by Gauss-Hermite series, which are able to properly reproduce the line profiles. The emission-line fluxes for the disc component are computed by the subtraction of the fluxes of the broad (outflow) component, obtained from the nuclear spectrum, from the fluxes measured in the integrated spectrum.

In the study of ionised outflows, one of the key properties to be considered is the electron density ($N_e$). We estimate $N_e$ for the outflow using the [S\,{\sc ii}]$\lambda \lambda$6716,6731 emission line fluxes measured for the outflow component, and assuming a constant electron temperature (T$_e$) of 10\,000 K \citep{Osterbrock2006}.
To calculate the electron density, we use the {\sc pyneb} \citep{PyNeb2015}, a Python package developed to analyse the emission line emissivities. We obtain an electron density for the ionised gas disc of $N_{e, \, {\rm disc}} = 6_{-2}^{+3} \times 10^2$  cm$^{-3}$, by subtracting the flux contribution of the outflow component. And for the outflow, we obtain a value of $N_{e, \, {\rm out}} = 6_{-3}^{+10} \times 10^3$  cm$^{-3}$.
The gas disc electron density is similar to the value estimated by  \cite{Ricci2014b}, $N_e \sim$ 874 cm$^{-3}$.
As we can see, the estimated density for the gas disc is within the expected range ($N_e \sim 10^2$ cm$^{-3}$) for narrow line regions (NLR) or H{\sc ii} Regions  \citep{Osterbrock2006,blandford2013active,dors15,freitas18,kakkad18} and is much smaller than that found for the outflow component. For the outflow density, a wide range of values have been derived and adopted in the literature: 
$10^2$ cm$^{-3}$ \citep{2014MNRAS.437.1708SchnorrMuller,2014MNRAS.441.3306Harrison,2017A&A...601A.143Fiore,Davies2020}, $10^3$ cm$^{-3}$ \citep{2016MNRAS.457..972SchnorrMuller,Davies2020} and $10^4 - 10^5$ cm$^{-3}$ \citep{2011MNRAS.410.1527Holt,2019MNRAS.486.4290Baron}. The [S\,{\sc ii}] method is only reliable within the limit of $N_e > 10^2$ cm$^{-3}$ and $N_e < 10^4$ cm$^{-3}$ 
\citep{Osterbrock2006,2019MNRAS.486.4290Baron,Davies2020}, therefore, our estimate for the outflow density is close to the limit in which it is still possible to use the $N_e$ estimation method from the [S\,{\sc ii}] doublet ratio, given the uncertainties in the measurements. 
However, $N_e$ obtained using the [S\,{\sc ii}] line ratio method generates significantly lower values than using other methods, such as using auroral and transauroral lines \citep{2011MNRAS.410.1527Holt} and the ionisation parameter \citep{2019MNRAS.486.4290Baron,Davies2020}. As discussed in \citet{Davies2020}, the interpretation of the lower values obtained from the [S\,{\sc ii}] doublet in AGN is that most of the [S\,{\sc ii}] emission arises in partially ionised zones, differently from the fully ionised region where higher ionisation lines are produced.  Thus, the $N_e$ value derived from the broad components of the [S\,{\sc ii}] lines, likely represents a lower limit to the outflow density.

To estimate the masses of ionised gas in the outflow and the gas disc, we use the extinction corrected H$\alpha$ luminosity ($L_{\rm H\alpha}$), estimated via colour excess by considering the case B recombination at an effective temperature of T$_e = 10\,000$K and the \cite{cardelli_1989} reddening law \citep{2019MNRAS.486.4290Baron,rogerio_2021}. 
We found $L_{{\rm H\alpha}, \,{\rm out}} = (2.4\pm0.2) \times 10^{40}$ erg s$^{-1}$ for the outflow component and $L_{{\rm H\alpha}, \, {\rm disc}}= (8.7\pm0.3)\times 10^{40}$ erg s$^{-1}$ for the gas disc component. We used these values to estimate the mass of ionised gas with the equation \citep{2020MNRAS.498.2632Hekatelyne}:

\begin{equation}\label{eq:mass}
    \frac{M}{\rm M_{\odot}} \approx 2.3 \times 10^{5} \frac{L_{41} (H_\alpha)}{n_{3}}, 
\end{equation}
where L$_{41}$(H$_\alpha$) is the H$\alpha$ luminosity in units of 10$^{41}$\,erg\,s$^{-1}$ and $n_3$ is the electron density ($N_e$) in units of 10$^3$\,cm$^{-3}$.
Using the $N_e$ values derived above, we obtain a value of $M_{\rm out} = (9.2 \pm 0.8) \times 10^3$ M$_\odot$ for the ionised gas mass in the outflow and $M_{\rm disc} = (3.3 \pm 0.1) \times 10^5$ M$_\odot$ for the ionised gas in the disc. These values represent an upper limit to the ionized gas mass for our galaxy, since we estimated a lower limit for the gas density by using the [S\,{\sc ii}] method, as discussed above.
Thus, the gas in the outflow corresponds to less than 3 per cent of the total amount of ionised gas in the inner 200 pc of NGC\,4546.

We can roughly estimate the mass outflow rate by assuming that the ionised gas is uniformly distributed within a sphere with a constant velocity and a given radius, as \citep[e.g.][]{2014MNRAS.441.3306Harrison,2017A&A...601A.143Fiore,2020A&A...642A.147Kakkad}:

\begin{equation}
    \dot{M}_{\rm out} \approx 3\frac{M_{\rm out}v_{\rm out}}{R_{\rm out}},
    \label{eq:out_rate}
\end{equation}
where $M_{\rm out}$ is the outflow mass, $v_{\rm out}$ is the outflow velocity and $R_{\rm out}$ is its radius. 
The outflow radius is assumed to be 0.35 arcsec ($\sim$ 30 pc), corresponding to the angular resolution of the GMOS data (0.7 arcsec FWHM).  As a proxy of the outflow velocity, we use the velocity dispersion of the broad H$\alpha$ component measured in the nuclear spectrum, of 320$\pm$30 km\,s$^{-1}$. The estimated mass outflow rate is $\dot{M}_{\rm out} = 0.3 \pm 0.1$ M$_\odot$ yr$^{-1}$. 

To quantify the impact of this outflow on the host galaxy, we also estimated the outflow kinetic power by:
\begin{equation}
    \dot{E}_{\rm out} \approx \frac{\dot{M}_{\rm out}}{2} v_{\rm out}^2, 
    \label{eq:kinetic_energy}
\end{equation}
where $\dot{M}$ is the outflow mass rate estimated above. 
We obtain an outflow kinetic power of $\dot{E}_{\rm out} = (9.7 \pm 2.9) \times 10^{39}$ erg s$^{-1}$
By comparing this quantity to the AGN bolometric luminosity, it is possible to investigate how efficient is the AGN feedback in suppressing star formation on the galaxy. We used the AGN bolometric luminosity estimated by \cite{Ricci2014b} for the NGC 4546 galaxy by using the relation between $L_{\rm bol}$ and [O {\sc iii}] luminosity ($L_{\rm bol}$\,/\,$L_{\rm [O \, III]} \sim 584$). \cite{Ricci2014b} found a value of order of $L_{\rm bol} \approx 8.7 \times 10^{42}$ erg s$^{−1}$ that gives us a kinetic efficiency (defined as the ratio between the kinetic power of the outflow and the AGN bolometric luminosity) of about 0.1 per cent. This value is much smaller than outflow coupling efficiencies required for AGN winds being efficient in suppressing star formation in the host galaxy, of at least  0.5 per cent \citep{Hopkins_Elvis_2010,Harrison_2018}. However, the total injected energy of the outflows in the host galaxy is not only represented by its kinetic power and a direct comparison between observed kinetic efficiency and the predicted coupling efficiency is not straightforward. Thus, even if the kinetic efficiency is lower than the limits predicted by models, the AGN feedback effect cannot be ruled out.



However, we have to be careful about the derived AGN feddback properties, since we are comparing kinematic measurements of one of the feedback gas phases with feedback models, that consider multi-gas phases of mechanical and radiative feedback modes \citep{Harrison_2018}.
Furthermore, many uncertainties are associated to the outflow properties, in special to the geometry and gas density. So, the quantities derived in this work can give us a clue about the order of the outflow properties that are important to understand the role that low luminosity AGN feedback play on the host galaxy. As mentioned before, AGN feedback is an important ingredient in models to explain galaxy formation and evolution, due to its effects on the host galaxy star formation \citep{2005Natur.433..604DiMatteo,fabian2012,Harrison_2018}.
Although low luminosity AGNs are not powerful enough to affect large-scale star formation in the host galaxy, it is important to understand how this process affects the innermost region.

\section{Conclusions}

We used GMOS-IFU observations of the inner 200 pc of NGC\,4546 to map the gas distribution and kinematics at an angular resolution of $\sim$60 pc. 
Our main conclusions are:

\begin{itemize}

    \item Using emission-line ratio diagnostic diagrams, we confirm that the nuclear emission of NGC\,4546 is consistent with emission of gas photoionised by an AGN. In circumnuclear regions, the BPT diagrams show mainly LINER-like emission, while the WHAN diagram shows mainly typical values of “Retired Galaxies” from spaxels at distances larger than $\sim 0.5$ arcsec from the nucleus. This indicates that the line emission from the circumnuclear region of NGC\,4546 may be produced by gas ionised by hot low-mass evolved stars.
    
    \item By modelling the ionised gas observed velocity field by a rotating disc, we found velocities of order of $\sim 60$ km s$^{-1}$ in the residual map (observed $-$ model circular velocity). These residual velocities may be due to remaining gas from a dwarf galaxy, previously captured by NGC\,4546, which is still not in orbital equilibrium with the galaxy, and likely triggered the nuclear activity in NGC\,4546.
    
    \item A broad component is necessary to model the emission-line profiles from the nuclear spectrum of NGC\,4546. This component is interpreted as being produced by an ionised outflow. We derive an electron density of $N_{e, \, {\rm out}} = 6_{-3}^{+10} \times 10^3$  cm$^{-3}$ for the outflow using the [S\,{\sc ii}] doublet and a mass of ionised gas of $M_{\rm out} = (9.2 \pm 0.8) \times 10^3$ M$_\odot$, corresponding to less than 3 per cent of the total mass of ionised gas in the inner 200 pc of NGC\,4546.

    \item Assuming a spherical outflow, we estimate a mass-outflow rate of $\dot{M}_{\rm out} = 0.3 \pm 0.1$ M$_\odot$ yr$^{-1}$ and a kinetic power of $\dot{E}_{\rm out} = (9.7 \pm 2.9) \times 10^{39}$ erg s$^{-1}$. The kinetic efficiency of the ionised outflow is about 0.1 per cent, lower than the theoretical predictions of coupling efficiencies to AGN winds become effective in suppressing the star formation in the host. However, it is worth mentioning that AGN winds are seen in multi-gas phases, and all phases should be took into account to compute the kinetic efficiency.

\end{itemize}

\section*{Acknowledgements}
We thank to an anonymous referee for the suggestions which helped us to improve this paper.
K.F.H. thanks the financial support from Coordena\c{c}\~{a}o de Aperfei\c{c}oamento de Pessoal de N\'ivel Superior - Brasil (CAPES) - Finance Code 001. T.V.R thanks Conselho Nacional de Desenvolvimento Cient\'ifico e Tecnol\'ogico (CNPq) for support under grant 306790/2019-0.  R.A.R. acknowledges the support from CNPq and Funda\c c\~ao de Amparo \`a pesquisa do Estado do Rio Grande do Sul. 
Based on observations obtained at the Gemini Observatory, which is operated by the Association of Universities for Research in Astronomy, Inc., under a cooperative agreement with the NSF on behalf of the Gemini partnership: the National Science Foundation (United States), National Research Council (Canada), CONICYT (Chile), Ministerio de Ciencia, Tecnolog\'{i}a e Innovaci\'{o}n Productiva (Argentina), Minist\'{e}rio da Ci\^{e}ncia, Tecnologia e Inova\c{c}\~{a}o (Brazil), and Korea Astronomy and Space Science Institute (Republic of Korea). This research has made use of NASA's Astrophysics Data System Bibliographic Services. This research has made use of the NASA/IPAC Extragalactic Database (NED), which is operated by the Jet Propulsion Laboratory, California Institute of Technology, under contract with the National Aeronautics and Space Administration.

\section*{Data Availability}
The data used in this work are publicly available online via the GEMINI archive https://archive.gemini.edu/searchform, under the program code GS-2008A-Q-51. 



\bibliographystyle{mnras}
\bibliography{paper_r1} 

\begin{thebibliography}{}
\makeatletter
\relax
\def\mn@urlcharsother{\let\do\@makeother \do\$\do\&\do\#\do\^\do\_\do\%\do\~}
\def\mn@doi{\begingroup\mn@urlcharsother \@ifnextchar [ {\mn@doi@}
  {\mn@doi@[]}}
\def\mn@doi@[#1]#2{\def\@tempa{#1}\ifx\@tempa\@empty \href
  {http://dx.doi.org/#2} {doi:#2}\else \href {http://dx.doi.org/#2} {#1}\fi
  \endgroup}
\def\mn@eprint#1#2{\mn@eprint@#1:#2::\@nil}
\def\mn@eprint@arXiv#1{\href {http://arxiv.org/abs/#1} {{\tt arXiv:#1}}}
\def\mn@eprint@dblp#1{\href {http://dblp.uni-trier.de/rec/bibtex/#1.xml}
  {dblp:#1}}
\def\mn@eprint@#1:#2:#3:#4\@nil{\def\@tempa {#1}\def\@tempb {#2}\def\@tempc
  {#3}\ifx \@tempc \@empty \let \@tempc \@tempb \let \@tempb \@tempa \fi \ifx
  \@tempb \@empty \def\@tempb {arXiv}\fi \@ifundefined
  {mn@eprint@\@tempb}{\@tempb:\@tempc}{\expandafter \expandafter \csname
  mn@eprint@\@tempb\endcsname \expandafter{\@tempc}}}

\bibitem[\protect\citeauthoryear{{Allington-Smith} et~al.,}{{Allington-Smith}
  et~al.}{2002}]{AllingtonSmith_2002}
{Allington-Smith} J.,  et~al., 2002, \mn@doi [\pasp] {10.1086/341712}, \href
  {https://ui.adsabs.harvard.edu/abs/2002PASP..114..892A} {114, 892}

\bibitem[\protect\citeauthoryear{{Baldwin}, {Phillips}  \&
  {Terlevich}}{{Baldwin} et~al.}{1981}]{bpt_1981}
{Baldwin} J.~A.,  {Phillips} M.~M.,   {Terlevich} R.,  1981, \mn@doi [\pasp]
  {10.1086/130766}, \href
  {https://ui.adsabs.harvard.edu/abs/1981PASP...93....5B} {93, 5}

\bibitem[\protect\citeauthoryear{{Baron} \& {Netzer}}{{Baron} \&
  {Netzer}}{2019}]{2019MNRAS.486.4290Baron}
{Baron} D.,  {Netzer} H.,  2019, \mn@doi [\mnras] {10.1093/mnras/stz1070},
  \href {https://ui.adsabs.harvard.edu/abs/2019MNRAS.486.4290B} {486, 4290}

\bibitem[\protect\citeauthoryear{{Belfiore} et~al.,}{{Belfiore}
  et~al.}{2016}]{2016MNRAS.461.3111Belfiore}
{Belfiore} F.,  et~al., 2016, \mn@doi [\mnras] {10.1093/mnras/stw1234}, \href
  {https://ui.adsabs.harvard.edu/abs/2016MNRAS.461.3111B} {461, 3111}

\bibitem[\protect\citeauthoryear{{Bettoni}, {Galletta}  \&
  {Oosterloo}}{{Bettoni} et~al.}{1991}]{Bettoni1991}
{Bettoni} D.,  {Galletta} G.,   {Oosterloo} T.,  1991, \mn@doi [\mnras]
  {10.1093/mnras/248.3.544}, \href
  {http://adsabs.harvard.edu/abs/1991MNRAS.248..544B} {248, 544}

\bibitem[\protect\citeauthoryear{{Binette}, {Magris}, {Stasi{\'n}ska}  \&
  {Bruzual}}{{Binette} et~al.}{1994}]{Binette_1994}
{Binette} L.,  {Magris} C.~G.,  {Stasi{\'n}ska} G.,   {Bruzual} A.~G.,  1994,
  \aap, \href {https://ui.adsabs.harvard.edu/abs/1994A&A...292...13B} {292, 13}

\bibitem[\protect\citeauthoryear{Blandford, Netzer, Woltjer, Courvoisier  \&
  Mayor}{Blandford et~al.}{2013}]{blandford2013active}
Blandford P.,  Netzer P.,  Woltjer P.,  Courvoisier T.,   Mayor P.,  2013,
  Active Galactic Nuclei.
Saas-Fee Advanced Course, Springer Berlin Heidelberg, \url
  {https://books.google.com.br/books?id=JpvsCAAAQBAJ}

\bibitem[\protect\citeauthoryear{{Brum} et~al.,}{{Brum}
  et~al.}{2019}]{2019MNRAS.486..691Brum}
{Brum} C.,  et~al., 2019, \mn@doi [\mnras] {10.1093/mnras/stz893}, \href
  {https://ui.adsabs.harvard.edu/abs/2019MNRAS.486..691B} {486, 691}

\bibitem[\protect\citeauthoryear{{Cappellari}}{{Cappellari}}{2017}]{Cappellari2017}
{Cappellari} M.,  2017, \mn@doi [\mnras] {10.1093/mnras/stw3020}, \href
  {http://adsabs.harvard.edu/abs/2017MNRAS.466..798C} {466, 798}

\bibitem[\protect\citeauthoryear{{Cardelli}, {Clayton}  \& {Mathis}}{{Cardelli}
  et~al.}{1989}]{cardelli_1989}
{Cardelli} J.~A.,  {Clayton} G.~C.,   {Mathis} J.~S.,  1989, \mn@doi [\apj]
  {10.1086/167900}, \href
  {https://ui.adsabs.harvard.edu/abs/1989ApJ...345..245C} {345, 245}

\bibitem[\protect\citeauthoryear{{Cattaneo} et~al.,}{{Cattaneo}
  et~al.}{2009}]{2009Cattaneo}
{Cattaneo} A.,  et~al., 2009, \mn@doi [\nat] {10.1038/nature08135}, \href
  {https://ui.adsabs.harvard.edu/abs/2009Natur.460..213C} {460, 213}

\bibitem[\protect\citeauthoryear{{Cid Fernandes}, {Stasi{\'n}ska},
  {Schlickmann}, {Mateus}, {Vale Asari}, {Schoenell}  \& {Sodr{\'e}}}{{Cid
  Fernandes} et~al.}{2010}]{Cid_Fernandes_2010}
{Cid Fernandes} R.,  {Stasi{\'n}ska} G.,  {Schlickmann} M.~S.,  {Mateus} A.,
  {Vale Asari} N.,  {Schoenell} W.,   {Sodr{\'e}} L.,  2010, \mn@doi [\mnras]
  {10.1111/j.1365-2966.2009.16185.x}, \href
  {https://ui.adsabs.harvard.edu/abs/2010MNRAS.403.1036C} {403, 1036}

\bibitem[\protect\citeauthoryear{{Cid Fernandes}, {Stasi{\'n}ska}, {Mateus}  \&
  {Vale Asari}}{{Cid Fernandes} et~al.}{2011}]{whan_cid_fernandes_2011}
{Cid Fernandes} R.,  {Stasi{\'n}ska} G.,  {Mateus} A.,   {Vale Asari} N.,
  2011, \mn@doi [\mnras] {10.1111/j.1365-2966.2011.18244.x}, \href
  {https://ui.adsabs.harvard.edu/abs/2011MNRAS.413.1687C} {413, 1687}

\bibitem[\protect\citeauthoryear{{Comer{\'o}n}, {Knapen}, {Ramos Almeida}  \&
  {Watkins}}{{Comer{\'o}n} et~al.}{2021}]{2021A&A...645A.130Comeron}
{Comer{\'o}n} S.,  {Knapen} J.~H.,  {Ramos Almeida} C.,   {Watkins} A.~E.,
  2021, \mn@doi [\aap] {10.1051/0004-6361/202039382}, \href
  {https://ui.adsabs.harvard.edu/abs/2021A&A...645A.130C} {645, A130}

\bibitem[\protect\citeauthoryear{{Davies} et~al.,}{{Davies}
  et~al.}{2014}]{2014ApJ...792..101Davies}
{Davies} R.~I.,  et~al., 2014, \mn@doi [\apj] {10.1088/0004-637X/792/2/101},
  \href {https://ui.adsabs.harvard.edu/abs/2014ApJ...792..101D} {792, 101}

\bibitem[\protect\citeauthoryear{{Davies} et~al.,}{{Davies}
  et~al.}{2020}]{Davies2020}
{Davies} R.,  et~al., 2020, \mn@doi [\mnras] {10.1093/mnras/staa2413}, \href
  {https://ui.adsabs.harvard.edu/abs/2020MNRAS.498.4150D} {498, 4150}

\bibitem[\protect\citeauthoryear{{Di Matteo}, {Springel}  \& {Hernquist}}{{Di
  Matteo} et~al.}{2005}]{2005Natur.433..604DiMatteo}
{Di Matteo} T.,  {Springel} V.,   {Hernquist} L.,  2005, \mn@doi [\nat]
  {10.1038/nature03335}, \href
  {https://ui.adsabs.harvard.edu/abs/2005Natur.433..604D} {433, 604}

\bibitem[\protect\citeauthoryear{{Diniz}, {Pastoriza}, {Hernandez-Jimenez},
  {Riffel}, {Ricci}, {Steiner}  \& {Riffel}}{{Diniz}
  et~al.}{2017}]{2017MNRAS.470.1703Diniz}
{Diniz} S. I.~F.,  {Pastoriza} M.~G.,  {Hernandez-Jimenez} J.~A.,  {Riffel} R.,
   {Ricci} T.~V.,  {Steiner} J.~E.,   {Riffel} R.~A.,  2017, \mn@doi [\mnras]
  {10.1093/mnras/stx1322}, \href
  {https://ui.adsabs.harvard.edu/abs/2017MNRAS.470.1703D} {470, 1703}

\bibitem[\protect\citeauthoryear{{Dors}, {Cardaci}, {H{\"a}gele}, {Rodrigues},
  {Grebel}, {Pilyugin}, {Freitas-Lemes}  \& {Krabbe}}{{Dors}
  et~al.}{2015}]{dors15}
{Dors} O.~L.,  {Cardaci} M.~V.,  {H{\"a}gele} G.~F.,  {Rodrigues} I.,  {Grebel}
  E.~K.,  {Pilyugin} L.~S.,  {Freitas-Lemes} P.,   {Krabbe} A.~C.,  2015,
  \mn@doi [\mnras] {10.1093/mnras/stv1916}, \href
  {https://ui.adsabs.harvard.edu/abs/2015MNRAS.453.4102D} {453, 4102}

\bibitem[\protect\citeauthoryear{{Eliche-Moral}, {Rodr{\'\i}guez-P{\'e}rez},
  {Borlaff}, {Querejeta}  \& {Tapia}}{{Eliche-Moral}
  et~al.}{2018}]{2018A&A...617A.113Eliche_Moral}
{Eliche-Moral} M.~C.,  {Rodr{\'\i}guez-P{\'e}rez} C.,  {Borlaff} A.,
  {Querejeta} M.,   {Tapia} T.,  2018, \mn@doi [\aap]
  {10.1051/0004-6361/201832911}, \href
  {https://ui.adsabs.harvard.edu/abs/2018A&A...617A.113E} {617, A113}

\bibitem[\protect\citeauthoryear{{Emonts}, {Colina}, {Piqueras-L{\'o}pez},
  {Garcia-Burillo}, {Pereira-Santaella}, {Arribas}, {Labiano}  \&
  {Alonso-Herrero}}{{Emonts} et~al.}{2017}]{emonts17}
{Emonts} B.~H.~C.,  {Colina} L.,  {Piqueras-L{\'o}pez} J.,  {Garcia-Burillo}
  S.,  {Pereira-Santaella} M.,  {Arribas} S.,  {Labiano} A.,   {Alonso-Herrero}
  A.,  2017, \mn@doi [\aap] {10.1051/0004-6361/201731508}, \href
  {https://ui.adsabs.harvard.edu/abs/2017A&A...607A.116E} {607, A116}

\bibitem[\protect\citeauthoryear{{Emsellem} et~al.,}{{Emsellem}
  et~al.}{2004}]{emsellem06_sauronIII}
{Emsellem} E.,  et~al., 2004, \mn@doi [\mnras]
  {10.1111/j.1365-2966.2004.07948.x}, \href
  {https://ui.adsabs.harvard.edu/abs/2004MNRAS.352..721E} {352, 721}

\bibitem[\protect\citeauthoryear{{Emsellem}, {Fathi}, {Wozniak}, {Ferruit},
  {Mundell}  \& {Schinnerer}}{{Emsellem} et~al.}{2006}]{emsellem06}
{Emsellem} E.,  {Fathi} K.,  {Wozniak} H.,  {Ferruit} P.,  {Mundell} C.~G.,
  {Schinnerer} E.,  2006, \mn@doi [\mnras] {10.1111/j.1365-2966.2005.09716.x},
  \href {https://ui.adsabs.harvard.edu/abs/2006MNRAS.365..367E} {365, 367}

\bibitem[\protect\citeauthoryear{{Escudero}, {Faifer}, {Smith Castelli},
  {Norris}  \& {Forte}}{{Escudero} et~al.}{2020}]{2020MNRAS.493.2253Escudero}
{Escudero} C.~G.,  {Faifer} F.~R.,  {Smith Castelli} A.~V.,  {Norris} M.~A.,
  {Forte} J.~C.,  2020, \mn@doi [\mnras] {10.1093/mnras/staa392}, \href
  {https://ui.adsabs.harvard.edu/abs/2020MNRAS.493.2253E} {493, 2253}

\bibitem[\protect\citeauthoryear{{Fabian}}{{Fabian}}{2012}]{fabian2012}
{Fabian} A.~C.,  2012, \mn@doi [\araa] {10.1146/annurev-astro-081811-125521},
  \href {https://ui.adsabs.harvard.edu/abs/2012ARA&A..50..455F} {50, 455}

\bibitem[\protect\citeauthoryear{{Falc{\'o}n-Barroso},
  {S{\'a}nchez-Bl{\'a}zquez}, {Vazdekis}, {Ricciardelli}, {Cardiel}, {Cenarro},
  {Gorgas}  \& {Peletier}}{{Falc{\'o}n-Barroso}
  et~al.}{2011}]{2011A&A...532A..95FalconBarroso}
{Falc{\'o}n-Barroso} J.,  {S{\'a}nchez-Bl{\'a}zquez} P.,  {Vazdekis} A.,
  {Ricciardelli} E.,  {Cardiel} N.,  {Cenarro} A.~J.,  {Gorgas} J.,
  {Peletier} R.~F.,  2011, \mn@doi [\aap] {10.1051/0004-6361/201116842}, \href
  {https://ui.adsabs.harvard.edu/abs/2011A&A...532A..95F} {532, A95}

\bibitem[\protect\citeauthoryear{{Ferland} \& {Netzer}}{{Ferland} \&
  {Netzer}}{1983}]{1983Ferland_Netzer}
{Ferland} G.~J.,  {Netzer} H.,  1983, \mn@doi [\apj] {10.1086/160577}, \href
  {https://ui.adsabs.harvard.edu/abs/1983ApJ...264..105F} {264, 105}

\bibitem[\protect\citeauthoryear{{Fiore} et~al.,}{{Fiore}
  et~al.}{2017}]{2017A&A...601A.143Fiore}
{Fiore} F.,  et~al., 2017, \mn@doi [\aap] {10.1051/0004-6361/201629478}, \href
  {https://ui.adsabs.harvard.edu/abs/2017A&A...601A.143F} {601, A143}

\bibitem[\protect\citeauthoryear{{Fischer}, {Crenshaw}, {Kraemer}, {Schmitt},
  {Storchi-Bergmann}  \& {Riffel}}{{Fischer}
  et~al.}{2015}]{2015ApJ...799..234Fischer}
{Fischer} T.~C.,  {Crenshaw} D.~M.,  {Kraemer} S.~B.,  {Schmitt} H.~R.,
  {Storchi-Bergmann} T.,   {Riffel} R.~A.,  2015, \mn@doi [\apj]
  {10.1088/0004-637X/799/2/234}, \href
  {https://ui.adsabs.harvard.edu/abs/2015ApJ...799..234F} {799, 234}

\bibitem[\protect\citeauthoryear{{Freitas} et~al.,}{{Freitas}
  et~al.}{2018}]{freitas18}
{Freitas} I.~C.,  et~al., 2018, \mn@doi [\mnras] {10.1093/mnras/sty303}, \href
  {https://ui.adsabs.harvard.edu/abs/2018MNRAS.476.2760F} {476, 2760}

\bibitem[\protect\citeauthoryear{{Gallagher}, {Maiolino}, {Belfiore}, {Drory},
  {Riffel}  \& {Riffel}}{{Gallagher} et~al.}{2019}]{gallagher19}
{Gallagher} R.,  {Maiolino} R.,  {Belfiore} F.,  {Drory} N.,  {Riffel} R.,
  {Riffel} R.~A.,  2019, \mn@doi [\mnras] {10.1093/mnras/stz564}, \href
  {https://ui.adsabs.harvard.edu/abs/2019MNRAS.485.3409G} {485, 3409}

\bibitem[\protect\citeauthoryear{{Galletta}}{{Galletta}}{1987}]{Galleta1987}
{Galletta} G.,  1987, \mn@doi [\apj] {10.1086/165389}, \href
  {http://adsabs.harvard.edu/abs/1987ApJ...318..531G} {318, 531}

\bibitem[\protect\citeauthoryear{{Gao}, {Ho}, {Barth}  \& {Li}}{{Gao}
  et~al.}{2018}]{2018ApJ.862.100Gao}
{Gao} H.,  {Ho} L.~C.,  {Barth} A.~J.,   {Li} Z.-Y.,  2018, \mn@doi [\apj]
  {10.3847/1538-4357/aacdac}, \href
  {https://ui.adsabs.harvard.edu/abs/2018ApJ...862..100G} {862, 100}

\bibitem[\protect\citeauthoryear{{Gerhard}}{{Gerhard}}{1993}]{1993Gerhard}
{Gerhard} O.~E.,  1993, \mn@doi [\mnras] {10.1093/mnras/265.1.213}, \href
  {https://ui.adsabs.harvard.edu/abs/1993MNRAS.265..213G} {265, 213}

\bibitem[\protect\citeauthoryear{{Greene}, {Zakamska}  \& {Smith}}{{Greene}
  et~al.}{2012}]{greene12}
{Greene} J.~E.,  {Zakamska} N.~L.,   {Smith} P.~S.,  2012, \mn@doi [\apj]
  {10.1088/0004-637X/746/1/86}, \href
  {https://ui.adsabs.harvard.edu/abs/2012ApJ...746...86G} {746, 86}

\bibitem[\protect\citeauthoryear{{Halpern} \& {Steiner}}{{Halpern} \&
  {Steiner}}{1983}]{1983Halpern_Steiner}
{Halpern} J.~P.,  {Steiner} J.~E.,  1983, \mn@doi [\apjl] {10.1086/184051},
  \href {https://ui.adsabs.harvard.edu/abs/1983ApJ...269L..37H} {269, L37}

\bibitem[\protect\citeauthoryear{{Harrison}}{{Harrison}}{2017}]{2017NatAs...1E.165Harrison}
{Harrison} C.~M.,  2017, \mn@doi [Nature Astronomy] {10.1038/s41550-017-0165},
  \href {https://ui.adsabs.harvard.edu/abs/2017NatAs...1E.165H} {1, 0165}

\bibitem[\protect\citeauthoryear{{Harrison}, {Alexander}, {Mullaney}  \&
  {Swinbank}}{{Harrison} et~al.}{2014}]{2014MNRAS.441.3306Harrison}
{Harrison} C.~M.,  {Alexander} D.~M.,  {Mullaney} J.~R.,   {Swinbank} A.~M.,
  2014, \mn@doi [\mnras] {10.1093/mnras/stu515}, \href
  {https://ui.adsabs.harvard.edu/abs/2014MNRAS.441.3306H} {441, 3306}

\bibitem[\protect\citeauthoryear{Harrison, Costa, Tadhunter, Flütsch, Kakkad,
  Perna  \& Vietri}{Harrison et~al.}{2018}]{Harrison_2018}
Harrison C.~M.,  Costa T.,  Tadhunter C.~N.,  Flütsch A.,  Kakkad D.,  Perna
  M.,   Vietri G.,  2018, \mn@doi [Nature Astronomy]
  {10.1038/s41550-018-0403-6}, 2, 198–205

\bibitem[\protect\citeauthoryear{{Heckman}}{{Heckman}}{1980}]{heckmann1980}
{Heckman} T.~M.,  1980, \aap, \href
  {https://ui.adsabs.harvard.edu/abs/1980A&A....87..152H} {87, 152}

\bibitem[\protect\citeauthoryear{{Hekatelyne}, {Riffel}, {Storchi-Bergmann},
  {Kharb}, {Robinson}, {Sales}  \& {Cassanta}}{{Hekatelyne}
  et~al.}{2020}]{2020MNRAS.498.2632Hekatelyne}
{Hekatelyne} C.,  {Riffel} R.~A.,  {Storchi-Bergmann} T.,  {Kharb} P.,
  {Robinson} A.,  {Sales} D.,   {Cassanta} C.~M.,  2020, \mn@doi [\mnras]
  {10.1093/mnras/staa2479}, \href
  {https://ui.adsabs.harvard.edu/abs/2020MNRAS.498.2632H} {498, 2632}

\bibitem[\protect\citeauthoryear{{Ho}}{{Ho}}{2008}]{Ho_2008}
{Ho} L.~C.,  2008, \mn@doi [\araa] {10.1146/annurev.astro.45.051806.110546},
  \href {https://ui.adsabs.harvard.edu/abs/2008ARA&A..46..475H} {46, 475}

\bibitem[\protect\citeauthoryear{{Ho}, {Filippenko}, {Sargent}  \& {Peng}}{{Ho}
  et~al.}{1997a}]{Ho_Filippenko_1997_IV}
{Ho} L.~C.,  {Filippenko} A.~V.,  {Sargent} W. L.~W.,   {Peng} C.~Y.,  1997a,
  \mn@doi [\apjs] {10.1086/313042}, \href
  {https://ui.adsabs.harvard.edu/abs/1997ApJS..112..391H} {112, 391}

\bibitem[\protect\citeauthoryear{{Ho}, {Filippenko}  \& {Sargent}}{{Ho}
  et~al.}{1997b}]{Ho_Filippenko1997_V}
{Ho} L.~C.,  {Filippenko} A.~V.,   {Sargent} W. L.~W.,  1997b, \mn@doi [\apj]
  {10.1086/304638}, \href
  {https://ui.adsabs.harvard.edu/abs/1997ApJ...487..568H} {487, 568}

\bibitem[\protect\citeauthoryear{{Holt}, {Tadhunter}, {Morganti}  \&
  {Emonts}}{{Holt} et~al.}{2011}]{2011MNRAS.410.1527Holt}
{Holt} J.,  {Tadhunter} C.~N.,  {Morganti} R.,   {Emonts} B.~H.~C.,  2011,
  \mn@doi [\mnras] {10.1111/j.1365-2966.2010.17535.x}, \href
  {https://ui.adsabs.harvard.edu/abs/2011MNRAS.410.1527H} {410, 1527}

\bibitem[\protect\citeauthoryear{{Hopkins} \& {Elvis}}{{Hopkins} \&
  {Elvis}}{2010}]{Hopkins_Elvis_2010}
{Hopkins} P.~F.,  {Elvis} M.,  2010, \mn@doi [\mnras]
  {10.1111/j.1365-2966.2009.15643.x}, \href
  {https://ui.adsabs.harvard.edu/abs/2010MNRAS.401....7H} {401, 7}

\bibitem[\protect\citeauthoryear{{Kakkad} et~al.,}{{Kakkad}
  et~al.}{2018}]{kakkad18}
{Kakkad} D.,  et~al., 2018, \mn@doi [\aap] {10.1051/0004-6361/201832790}, \href
  {https://ui.adsabs.harvard.edu/abs/2018A&A...618A...6K} {618, A6}

\bibitem[\protect\citeauthoryear{{Kakkad} et~al.,}{{Kakkad}
  et~al.}{2020}]{2020A&A...642A.147Kakkad}
{Kakkad} D.,  et~al., 2020, \mn@doi [\aap] {10.1051/0004-6361/202038551}, \href
  {https://ui.adsabs.harvard.edu/abs/2020A&A...642A.147K} {642, A147}

\bibitem[\protect\citeauthoryear{{Kauffmann} et~al.,}{{Kauffmann}
  et~al.}{2003}]{kauffmann_2003}
{Kauffmann} G.,  et~al., 2003, \mn@doi [\mnras]
  {10.1111/j.1365-2966.2003.07154.x}, \href
  {https://ui.adsabs.harvard.edu/abs/2003MNRAS.346.1055K} {346, 1055}

\bibitem[\protect\citeauthoryear{{Kewley}, {Dopita}, {Sutherland}, {Heisler}
  \& {Trevena}}{{Kewley} et~al.}{2001}]{kewley2001b}
{Kewley} L.~J.,  {Dopita} M.~A.,  {Sutherland} R.~S.,  {Heisler} C.~A.,
  {Trevena} J.,  2001, \mn@doi [\apj] {10.1086/321545}, \href
  {https://ui.adsabs.harvard.edu/abs/2001ApJ...556..121K} {556, 121}

\bibitem[\protect\citeauthoryear{{Kewley}, {Groves}, {Kauffmann}  \&
  {Heckman}}{{Kewley} et~al.}{2006}]{Kewley_2006}
{Kewley} L.~J.,  {Groves} B.,  {Kauffmann} G.,   {Heckman} T.,  2006, \mn@doi
  [\mnras] {10.1111/j.1365-2966.2006.10859.x}, \href
  {https://ui.adsabs.harvard.edu/abs/2006MNRAS.372..961K} {372, 961}

\bibitem[\protect\citeauthoryear{{Kim}}{{Kim}}{1989}]{1989ApJ...346..653Kim}
{Kim} D.-W.,  1989, \mn@doi [\apj] {10.1086/168048}, \href
  {https://ui.adsabs.harvard.edu/abs/1989ApJ...346..653K} {346, 653}

\bibitem[\protect\citeauthoryear{{Kormendy} \& {Ho}}{{Kormendy} \&
  {Ho}}{2013}]{2013ARA&A..51..511Kormendy}
{Kormendy} J.,  {Ho} L.~C.,  2013, \mn@doi [\araa]
  {10.1146/annurev-astro-082708-101811}, \href
  {https://ui.adsabs.harvard.edu/abs/2013ARA&A..51..511K} {51, 511}

\bibitem[\protect\citeauthoryear{{Krajnovi{\'c}}, {Cappellari}, {de Zeeuw}  \&
  {Copin}}{{Krajnovi{\'c}} et~al.}{2006}]{Krajnovic2006_kinemetry}
{Krajnovi{\'c}} D.,  {Cappellari} M.,  {de Zeeuw} P.~T.,   {Copin} Y.,  2006,
  \mn@doi [\mnras] {10.1111/j.1365-2966.2005.09902.x}, \href
  {https://ui.adsabs.harvard.edu/abs/2006MNRAS.366..787K} {366, 787}

\bibitem[\protect\citeauthoryear{{Luridiana}, {Morisset}  \&
  {Shaw}}{{Luridiana} et~al.}{2015}]{PyNeb2015}
{Luridiana} V.,  {Morisset} C.,   {Shaw} R.~A.,  2015, \mn@doi [\aap]
  {10.1051/0004-6361/201323152}, \href
  {https://ui.adsabs.harvard.edu/abs/2015A&A...573A..42L} {573, A42}

\bibitem[\protect\citeauthoryear{{Magorrian} et~al.,}{{Magorrian}
  et~al.}{1998}]{magorrian1998}
{Magorrian} J.,  et~al., 1998, \mn@doi [\aj] {10.1086/300353}, \href
  {https://ui.adsabs.harvard.edu/abs/1998AJ....115.2285M} {115, 2285}

\bibitem[\protect\citeauthoryear{{Menezes}, {Ricci}, {Steiner}, {da Silva},
  {Ferrari}  \& {Borges}}{{Menezes} et~al.}{2019}]{menezes2019}
{Menezes} R.~B.,  {Ricci} T.~V.,  {Steiner} J.~E.,  {da Silva} P.,  {Ferrari}
  F.,   {Borges} B.~W.,  2019, \mn@doi [\mnras] {10.1093/mnras/sty3337}, \href
  {http://adsabs.harvard.edu/abs/2019MNRAS.483.3700M} {483, 3700}

\bibitem[\protect\citeauthoryear{{Molina}, {Eracleous}, {Barth}, {Maoz},
  {Runnoe}, {Ho}, {Shields}  \& {Walsh}}{{Molina} et~al.}{2018}]{Molina_2018}
{Molina} M.,  {Eracleous} M.,  {Barth} A.~J.,  {Maoz} D.,  {Runnoe} J.~C.,
  {Ho} L.~C.,  {Shields} J.~C.,   {Walsh} J.~L.,  2018, \mn@doi [\apj]
  {10.3847/1538-4357/aad5ed}, \href
  {https://ui.adsabs.harvard.edu/abs/2018ApJ...864...90M} {864, 90}

\bibitem[\protect\citeauthoryear{{Norris} \& {Kannappan}}{{Norris} \&
  {Kannappan}}{2011a}]{2011EAS....48..259Norris&Kannappan}
{Norris} M.~A.,  {Kannappan} S.~J.,  2011a, in {Koleva} M.,  {Prugniel} P.,
  {Vauglin} I.,  eds,  EAS Publications Series Vol. 48, EAS Publications
  Series. pp 259--260 (\mn@eprint {arXiv} {1009.2489}),
  \mn@doi{10.1051/eas/1148059}

\bibitem[\protect\citeauthoryear{{Norris} \& {Kannappan}}{{Norris} \&
  {Kannappan}}{2011b}]{2011MNRAS.414..739Noris&Kannappan}
{Norris} M.~A.,  {Kannappan} S.~J.,  2011b, \mn@doi [\mnras]
  {10.1111/j.1365-2966.2011.18440.x}, \href
  {https://ui.adsabs.harvard.edu/abs/2011MNRAS.414..739N} {414, 739}

\bibitem[\protect\citeauthoryear{{Norris}, {Escudero}, {Faifer}, {Kannappan},
  {Forte}  \& {van den Bosch}}{{Norris}
  et~al.}{2015}]{2015MNRAS.451.3615Norris}
{Norris} M.~A.,  {Escudero} C.~G.,  {Faifer} F.~R.,  {Kannappan} S.~J.,
  {Forte} J.~C.,   {van den Bosch} R. C.~E.,  2015, \mn@doi [\mnras]
  {10.1093/mnras/stv1221}, \href
  {https://ui.adsabs.harvard.edu/abs/2015MNRAS.451.3615N} {451, 3615}

\bibitem[\protect\citeauthoryear{{Nyland} et~al.,}{{Nyland}
  et~al.}{2016}]{Nyland_2016}
{Nyland} K.,  et~al., 2016, \mn@doi [\mnras] {10.1093/mnras/stw391}, \href
  {https://ui.adsabs.harvard.edu/abs/2016MNRAS.458.2221N} {458, 2221}

\bibitem[\protect\citeauthoryear{{Osterbrock} \& {Ferland}}{{Osterbrock} \&
  {Ferland}}{2006}]{Osterbrock2006}
{Osterbrock} D.~E.,  {Ferland} G.~J.,  2006, {Astrophysics of gaseous nebulae
  and active galactic nuclei}.
CA: University Science Books

\bibitem[\protect\citeauthoryear{{Perrotta}, {Hamann}, {Zakamska},
  {Alexandroff}, {Rupke}  \& {Wylezalek}}{{Perrotta} et~al.}{2019}]{perrota19}
{Perrotta} S.,  {Hamann} F.,  {Zakamska} N.~L.,  {Alexandroff} R.~M.,  {Rupke}
  D.,   {Wylezalek} D.,  2019, \mn@doi [\mnras] {10.1093/mnras/stz1993}, \href
  {https://ui.adsabs.harvard.edu/abs/2019MNRAS.488.4126P} {488, 4126}

\bibitem[\protect\citeauthoryear{{Phillips}, {Jenkins}, {Dopita}, {Sadler}  \&
  {Binette}}{{Phillips} et~al.}{1986}]{1986AJ.....91.1062Phillips}
{Phillips} M.~M.,  {Jenkins} C.~R.,  {Dopita} M.~A.,  {Sadler} E.~M.,
  {Binette} L.,  1986, \mn@doi [\aj] {10.1086/114083}, \href
  {https://ui.adsabs.harvard.edu/abs/1986AJ.....91.1062P} {91, 1062}

\bibitem[\protect\citeauthoryear{{Raimundo}}{{Raimundo}}{2021}]{2021A&A...650A..34Raimundo}
{Raimundo} S.~I.,  2021, \mn@doi [\aap] {10.1051/0004-6361/202040248}, \href
  {https://ui.adsabs.harvard.edu/abs/2021A&A...650A..34R} {650, A34}

\bibitem[\protect\citeauthoryear{{Raimundo}, {Davies}, {Canning}, {Celotti},
  {Fabian}  \& {Gandhi}}{{Raimundo} et~al.}{2017}]{2017MNRAS.464.4227Raimundo}
{Raimundo} S.~I.,  {Davies} R.~I.,  {Canning} R.~E.~A.,  {Celotti} A.,
  {Fabian} A.~C.,   {Gandhi} P.,  2017, \mn@doi [\mnras]
  {10.1093/mnras/stw2635}, \href
  {https://ui.adsabs.harvard.edu/abs/2017MNRAS.464.4227R} {464, 4227}

\bibitem[\protect\citeauthoryear{{Ricci} \& {Steiner}}{{Ricci} \&
  {Steiner}}{2020}]{Ricci2020}
{Ricci} T.~V.,  {Steiner} J.~E.,  2020, \mn@doi [\mnras]
  {10.1093/mnras/staa1398}, \href
  {https://ui.adsabs.harvard.edu/abs/2020MNRAS.495.2620R} {495, 2620}

\bibitem[\protect\citeauthoryear{{Ricci}, {Steiner}  \& {Menezes}}{{Ricci}
  et~al.}{2014a}]{Ricci2014a}
{Ricci} T.~V.,  {Steiner} J.~E.,   {Menezes} R.~B.,  2014a, \mn@doi [\mnras]
  {10.1093/mnras/stu441}, \href
  {http://adsabs.harvard.edu/abs/2014MNRAS.440.2419R} {440, 2419}

\bibitem[\protect\citeauthoryear{{Ricci}, {Steiner}  \& {Menezes}}{{Ricci}
  et~al.}{2014b}]{Ricci2014b}
{Ricci} T.~V.,  {Steiner} J.~E.,   {Menezes} R.~B.,  2014b, \mn@doi [\mnras]
  {10.1093/mnras/stu442}, \href
  {http://adsabs.harvard.edu/abs/2014MNRAS.440.2442R} {440, 2442}

\bibitem[\protect\citeauthoryear{{Ricci}, {Steiner}  \& {Menezes}}{{Ricci}
  et~al.}{2015}]{Ricci2015_III}
{Ricci} T.~V.,  {Steiner} J.~E.,   {Menezes} R.~B.,  2015, \mn@doi [\mnras]
  {10.1093/mnras/stv1156}, \href
  {http://adsabs.harvard.edu/abs/2015MNRAS.451.3728R} {451, 3728}

\bibitem[\protect\citeauthoryear{{Ricci}, {Steiner}  \& {Menezes}}{{Ricci}
  et~al.}{2016}]{Ricci2016}
{Ricci} T.~V.,  {Steiner} J.~E.,   {Menezes} R.~B.,  2016, \mn@doi [\mnras]
  {10.1093/mnras/stw2318}, \href
  {http://adsabs.harvard.edu/abs/2016MNRAS.463.3860R} {463, 3860}

\bibitem[\protect\citeauthoryear{{Riffel}}{{Riffel}}{2010}]{Riffel2010}
{Riffel} R.~A.,  2010, \mn@doi [\apss] {10.1007/s10509-010-0317-y}, \href
  {http://adsabs.harvard.edu/abs/2010Ap\%26SS.327..239R} {327, 239}

\bibitem[\protect\citeauthoryear{{Riffel}, {Storchi-Bergmann}  \&
  {Riffel}}{{Riffel} et~al.}{2015}]{2015MNRAS.451.3587Riffel}
{Riffel} R.~A.,  {Storchi-Bergmann} T.,   {Riffel} R.,  2015, \mn@doi [\mnras]
  {10.1093/mnras/stv1129}, \href
  {https://ui.adsabs.harvard.edu/abs/2015MNRAS.451.3587R} {451, 3587}

\bibitem[\protect\citeauthoryear{{Riffel}, {Storchi-Bergmann}, {Riffel},
  {Dahmer-Hahn}, {Diniz}, {Sch{\"o}nell}  \& {Dametto}}{{Riffel}
  et~al.}{2017}]{rogemar17_stellar}
{Riffel} R.~A.,  {Storchi-Bergmann} T.,  {Riffel} R.,  {Dahmer-Hahn} L.~G.,
  {Diniz} M.~R.,  {Sch{\"o}nell} A.~J.,   {Dametto} N.~Z.,  2017, \mn@doi
  [\mnras] {10.1093/mnras/stx1308}, \href
  {https://ui.adsabs.harvard.edu/abs/2017MNRAS.470..992R} {470, 992}

\bibitem[\protect\citeauthoryear{{Riffel}, {Storchi-Bergmann}, {Zakamska}  \&
  {Riffel}}{{Riffel} et~al.}{2020}]{rogemar20_n1275}
{Riffel} R.~A.,  {Storchi-Bergmann} T.,  {Zakamska} N.~L.,   {Riffel} R.,
  2020, \mn@doi [\mnras] {10.1093/mnras/staa1922}, \href
  {https://ui.adsabs.harvard.edu/abs/2020MNRAS.496.4857R} {496, 4857}

\bibitem[\protect\citeauthoryear{{Riffel} et~al.,}{{Riffel}
  et~al.}{2021}]{rogerio_2021}
{Riffel} R.,  et~al., 2021, \mn@doi [\mnras] {10.1093/mnras/staa3907}, \href
  {https://ui.adsabs.harvard.edu/abs/2021MNRAS.501.4064R} {501, 4064}

\bibitem[\protect\citeauthoryear{Ruschel-Dutra \& de Oliveira}{Ruschel-Dutra \&
  de~Oliveira}{2020}]{daniel_ruschel_dutra_2020_3945237}
Ruschel-Dutra D.,  de Oliveira B.~D.,  2020, danielrd6/ifscube v1.0,
  \mn@doi{10.5281/zenodo.3945237}

\bibitem[\protect\citeauthoryear{{Ruschel-Dutra} et~al.,}{{Ruschel-Dutra}
  et~al.}{2021}]{ruschel-dutra21}
{Ruschel-Dutra} D.,  et~al., 2021, \mn@doi [\mnras] {10.1093/mnras/stab2058},
  \href {https://ui.adsabs.harvard.edu/abs/2021MNRAS.507...74R} {507, 74}

\bibitem[\protect\citeauthoryear{{Sadler}, {Jenkins}  \& {Kotanyi}}{{Sadler}
  et~al.}{1989}]{1989MNRAS.240..591Sadler}
{Sadler} E.~M.,  {Jenkins} C.~R.,   {Kotanyi} C.~G.,  1989, \mn@doi [\mnras]
  {10.1093/mnras/240.3.591}, \href
  {https://ui.adsabs.harvard.edu/abs/1989MNRAS.240..591S} {240, 591}

\bibitem[\protect\citeauthoryear{{Sage} \& {Galletta}}{{Sage} \&
  {Galletta}}{1994}]{Sage1994}
{Sage} L.~J.,  {Galletta} G.,  1994, \mn@doi [\aj] {10.1086/117184}, \href
  {http://adsabs.harvard.edu/abs/1994AJ....108.1633S} {108, 1633}

\bibitem[\protect\citeauthoryear{{S{\'a}nchez-Bl{\'a}zquez}
  et~al.,}{{S{\'a}nchez-Bl{\'a}zquez}
  et~al.}{2006}]{2006MNRAS.371..703SanchezBlazquez}
{S{\'a}nchez-Bl{\'a}zquez} P.,  et~al., 2006, \mn@doi [\mnras]
  {10.1111/j.1365-2966.2006.10699.x}, \href
  {https://ui.adsabs.harvard.edu/abs/2006MNRAS.371..703S} {371, 703}

\bibitem[\protect\citeauthoryear{{Sarzi} et~al.,}{{Sarzi}
  et~al.}{2006}]{sarzi2006}
{Sarzi} M.,  et~al., 2006, \mn@doi [\mnras] {10.1111/j.1365-2966.2005.09839.x},
  \href {https://ui.adsabs.harvard.edu/abs/2006MNRAS.366.1151S} {366, 1151}

\bibitem[\protect\citeauthoryear{{Sarzi} et~al.,}{{Sarzi}
  et~al.}{2010}]{sarzi_2010}
{Sarzi} M.,  et~al., 2010, \mn@doi [\mnras] {10.1111/j.1365-2966.2009.16039.x},
  \href {https://ui.adsabs.harvard.edu/abs/2010MNRAS.402.2187S} {402, 2187}

\bibitem[\protect\citeauthoryear{{Schnorr-M{\"u}ller}, {Storchi-Bergmann},
  {Nagar}, {Robinson}, {Lena}, {Riffel}  \& {Couto}}{{Schnorr-M{\"u}ller}
  et~al.}{2014}]{2014MNRAS.437.1708SchnorrMuller}
{Schnorr-M{\"u}ller} A.,  {Storchi-Bergmann} T.,  {Nagar} N.~M.,  {Robinson}
  A.,  {Lena} D.,  {Riffel} R.~A.,   {Couto} G.~S.,  2014, \mn@doi [\mnras]
  {10.1093/mnras/stt2001}, \href
  {https://ui.adsabs.harvard.edu/abs/2014MNRAS.437.1708S} {437, 1708}

\bibitem[\protect\citeauthoryear{{Schnorr-M{\"u}ller}, {Storchi-Bergmann},
  {Robinson}, {Lena}  \& {Nagar}}{{Schnorr-M{\"u}ller}
  et~al.}{2016}]{2016MNRAS.457..972SchnorrMuller}
{Schnorr-M{\"u}ller} A.,  {Storchi-Bergmann} T.,  {Robinson} A.,  {Lena} D.,
  {Nagar} N.~M.,  2016, \mn@doi [\mnras] {10.1093/mnras/stw037}, \href
  {https://ui.adsabs.harvard.edu/abs/2016MNRAS.457..972S} {457, 972}

\bibitem[\protect\citeauthoryear{{She}, {Ho}  \& {Feng}}{{She}
  et~al.}{2017}]{2017ApJ...835..223She}
{She} R.,  {Ho} L.~C.,   {Feng} H.,  2017, \mn@doi [\apj]
  {10.3847/1538-4357/835/2/223}, \href
  {https://ui.adsabs.harvard.edu/abs/2017ApJ...835..223S} {835, 223}

\bibitem[\protect\citeauthoryear{{Singh} et~al.,}{{Singh}
  et~al.}{2013}]{2013A&A...558A..43Singh}
{Singh} R.,  et~al., 2013, \mn@doi [\aap] {10.1051/0004-6361/201322062}, \href
  {https://ui.adsabs.harvard.edu/abs/2013A&A...558A..43S} {558, A43}

\bibitem[\protect\citeauthoryear{{Slee}, {Sadler}, {Reynolds}  \&
  {Ekers}}{{Slee} et~al.}{1994}]{1994MNRAS.269..928Slee}
{Slee} O.~B.,  {Sadler} E.~M.,  {Reynolds} J.~E.,   {Ekers} R.~D.,  1994,
  \mn@doi [\mnras] {10.1093/mnras/269.4.928}, \href
  {https://ui.adsabs.harvard.edu/abs/1994MNRAS.269..928S} {269, 928}

\bibitem[\protect\citeauthoryear{{Stasi{\'n}ska} et~al.,}{{Stasi{\'n}ska}
  et~al.}{2008}]{Stasinska_2008}
{Stasi{\'n}ska} G.,  et~al., 2008, \mn@doi [\mnras]
  {10.1111/j.1745-3933.2008.00550.x}, \href
  {https://ui.adsabs.harvard.edu/abs/2008MNRAS.391L..29S} {391, L29}

\bibitem[\protect\citeauthoryear{{Van Dokkum}}{{Van
  Dokkum}}{2001}]{van_Dokkum_2001}
{Van Dokkum} P.~G.,  2001, \mn@doi [\pasp] {10.1086/323894}, \href
  {https://ui.adsabs.harvard.edu/abs/2001PASP..113.1420V} {113, 1420}

\bibitem[\protect\citeauthoryear{{Van de Voort}, {Davis}, {Kere{\v{s}}},
  {Quataert}, {Faucher-Gigu{\`e}re}  \& {Hopkins}}{{Van de Voort}
  et~al.}{2015}]{2015MNRAS.451.3269VandeVoort}
{Van de Voort} F.,  {Davis} T.~A.,  {Kere{\v{s}}} D.,  {Quataert} E.,
  {Faucher-Gigu{\`e}re} C.-A.,   {Hopkins} P.~F.,  2015, \mn@doi [\mnras]
  {10.1093/mnras/stv1217}, \href
  {https://ui.adsabs.harvard.edu/abs/2015MNRAS.451.3269V} {451, 3269}

\bibitem[\protect\citeauthoryear{{Van der Marel} \& {Franx}}{{Van der Marel} \&
  {Franx}}{1993}]{1993van_der_Marel_Franx}
{Van der Marel} R.~P.,  {Franx} M.,  1993, \mn@doi [\apj] {10.1086/172534},
  \href {https://ui.adsabs.harvard.edu/abs/1993ApJ...407..525V} {407, 525}

\bibitem[\protect\citeauthoryear{{Veilleux} \& {Osterbrock}}{{Veilleux} \&
  {Osterbrock}}{1987}]{veilleux_osterbrock_1987}
{Veilleux} S.,  {Osterbrock} D.~E.,  1987, \mn@doi [\apjs] {10.1086/191166},
  \href {https://ui.adsabs.harvard.edu/abs/1987ApJS...63..295V} {63, 295}

\bibitem[\protect\citeauthoryear{{Westfall} et~al.,}{{Westfall}
  et~al.}{2019}]{Westfall_2019}
{Westfall} K.~B.,  et~al., 2019, \mn@doi [\aj] {10.3847/1538-3881/ab44a2},
  \href {https://ui.adsabs.harvard.edu/abs/2019AJ....158..231W} {158, 231}

\bibitem[\protect\citeauthoryear{{Yan} \& {Blanton}}{{Yan} \&
  {Blanton}}{2012}]{2012ApJ...747...61Yan_Blanton}
{Yan} R.,  {Blanton} M.~R.,  2012, \mn@doi [\apj] {10.1088/0004-637X/747/1/61},
  \href {https://ui.adsabs.harvard.edu/abs/2012ApJ...747...61Y} {747, 61}

\bibitem[\protect\citeauthoryear{{Zubovas}, {Nayakshin}, {King}  \&
  {Wilkinson}}{{Zubovas} et~al.}{2013}]{2013MNRAS.433.3079Zubovas}
{Zubovas} K.,  {Nayakshin} S.,  {King} A.,   {Wilkinson} M.,  2013, \mn@doi
  [\mnras] {10.1093/mnras/stt952}, \href
  {https://ui.adsabs.harvard.edu/abs/2013MNRAS.433.3079Z} {433, 3079}

\makeatother
\end{thebibliography}




\appendix

\bsp	
\label{lastpage}
\end{document}